\documentclass[%
reprint,
superscriptaddress,
amsmath,amssymb
aps,
twocolumn,epsf,prl,
]{revtex4-2}

\usepackage{graphicx}
\usepackage{dcolumn}
\usepackage{bm}
\usepackage{amsmath}
\usepackage{amsfonts}
\usepackage{amssymb}
\usepackage[dvipsnames]{xcolor}
\usepackage{hyperref}
\usepackage{pbox}
\usepackage{pinlabel}

\begin{document}
	
	\title{Resistivity Exponents in 3D-Dirac Semimetals From Electron-Electron Interaction}
	
	\author{Niklas Wagner}
	\affiliation{Institut f\"ur Theoretische Physik und Astrophysik, Universit\"at W\"urzburg, 97074 W\"urzburg, Germany}
	\author{Sergio Ciuchi}%
	\affiliation{Dipartimento di Scienze Fisiche e Chimiche, Universit`a dell’Aquila, and Istituto dei Sistemi Complessi, CNR, Coppito-L’Aquila, Italy}
	\author{Alessandro Toschi}
	\affiliation{Institute of Solid State Physics, TU Wien, 1040 Vienna, Austria}
	\author{Björn Trauzettel}
	\affiliation{Institut f\"ur Theoretische Physik und Astrophysik and W\"urzburg-Dresden Cluster of Excellence ct.qmat, Universit\"at W\"urzburg, 97074 W\"urzburg, Germany}
	\author{Giorgio Sangiovanni}
	\affiliation{Institut f\"ur Theoretische Physik und Astrophysik and W\"urzburg-Dresden Cluster of Excellence ct.qmat, Universit\"at W\"urzburg, 97074 W\"urzburg, Germany}

\begin{abstract}
We study the resistivity of three-dimensional semimetals with linear dispersion in the presence of on-site electron-electron interaction.
The well-known quadratic temperature dependence of the resistivity of conventional metals is turned into an unusual $T^6$-behavior. 
An analogous change affects the thermal transport, preserving the linearity in $T$ of the ratio between thermal and electrical conductivities. These results hold from weak coupling up to the non-perturbative region of the Mott transition.
Our findings yield a natural explanation for the hitherto not understood large exponents characterizing the temperature-dependence of transport experiments on various topological semimetals.
\end{abstract}

	\maketitle

\noindent	
{\it Introduction} -- Topologically protected nodal semimetals are characterized by a linear energy-momentum relation and can be viewed as a condensed-matter realization of Dirac and Weyl high-energy particles \cite{armitage_weyl_2018,wehling_dirac_2014}.  
These materials are characterized by peculiar transport properties, as a result of their non-trivial electronic bandstructure and of conducting boundary modes. One of the most remarkable phenomena is the negative magnetoresistance of Weyl semimetals, a manifestation of the chiral anomaly  \cite{son_chiral_2013,ashby_chiral_2014}.
The impact of impurity scattering on the conductivity of three-dimensional (3D) Dirac semimetals has attracted a lot of attention: a residual conductivity is found for short-range random potentials \cite{tabert_optical_2016} but, at the same time, some of the universal transport features characterizing graphene emerge \cite{burkov_topological_2011,hosur_charge_2012}. 
While for small disorder, propagating Dirac fermions determine the transport properties, localized states appear in the opposite limit giving rise to the non-Anderson scenario \cite{syzranov_high-dimensional_2018}. 

\begin{figure}[t]
    \labellist
    \small\hair 2pt
    \pinlabel a) at 1 900
    \pinlabel b) at 1200 900
    \pinlabel c) at 1200 550
    \endlabellist
	\begin{minipage}{.5\linewidth}
       	\includegraphics[width=1.0\linewidth]{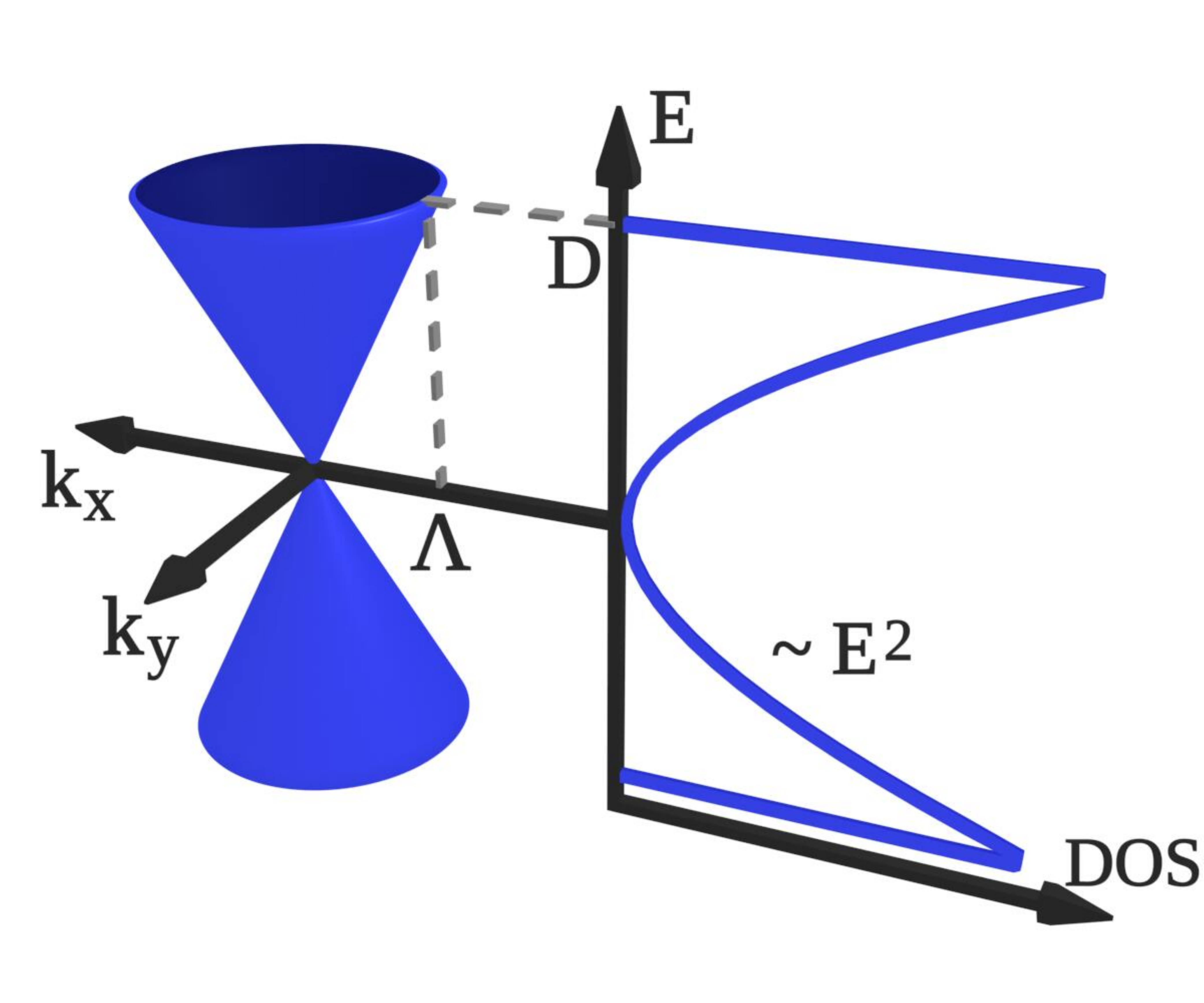}
	\end{minipage}
	\begin{minipage}{.4\linewidth}
    	\setlength\tabcolsep{1.3mm}
    	\def\arraystretch{2.4}
    	\setlength{\arrayrulewidth}{0.5mm}
        \centering
        \includegraphics[width=0.7\linewidth]{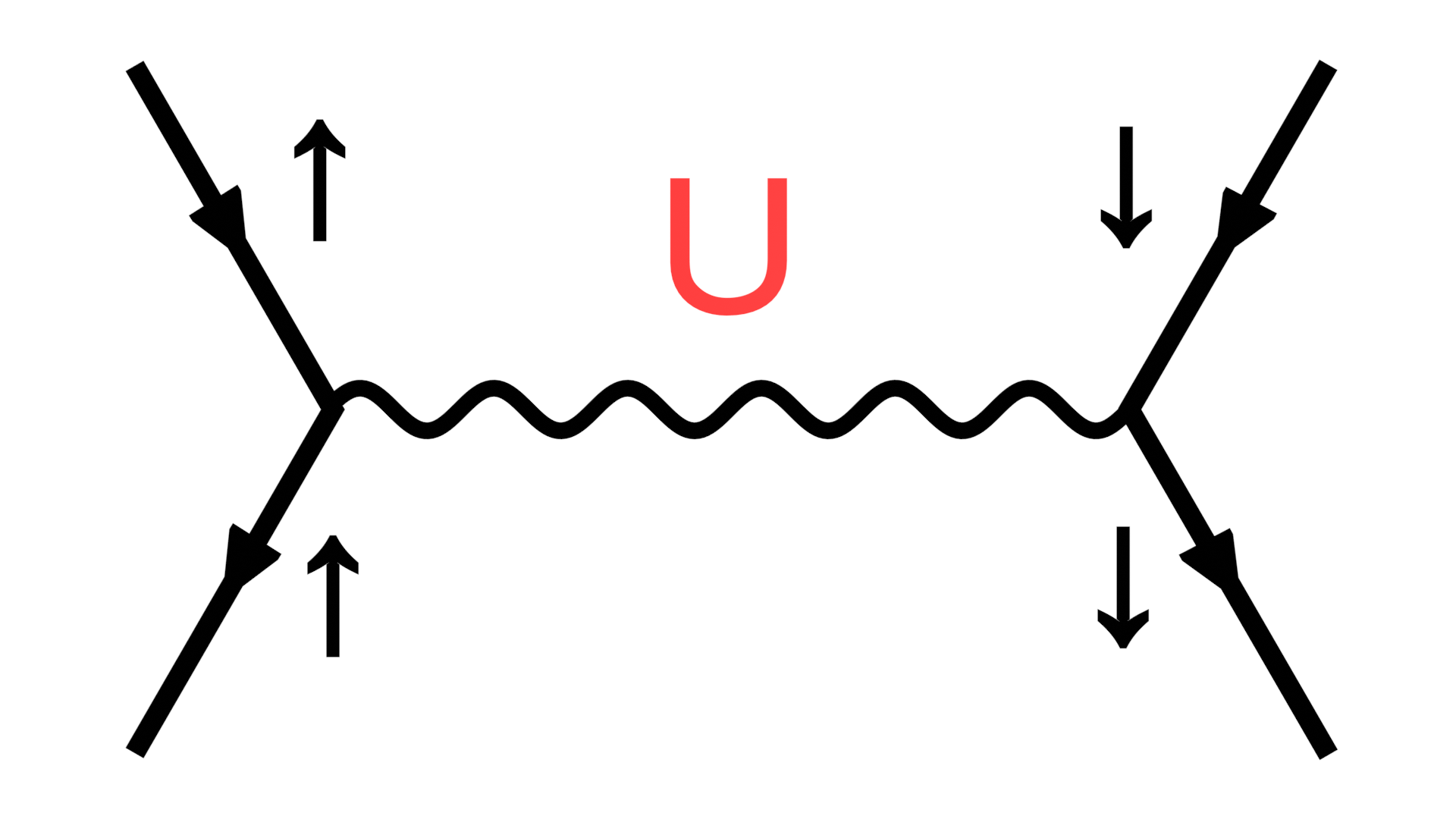}
    	\begin{tabular}{c|c|c}
        	  &\pbox{1cm}{Fermi\\liquid}& \pbox{1cm}{Dirac/\\Weyl} \\ \hline
        	\pbox{1.5 cm}{scattering\\rate} & \large$T^2$& \large$T^8$  \\
        	\hline
        	resistivity &\large $T^2$& \large$T^6$  \\ 
        \end{tabular}
	\end{minipage}
	\hspace{0.3cm}
	    \caption{
	    (a) Dirac/Weyl dispersion (for $k_z\!=\!0$) and DOS with cutoff $\Lambda$ in $k$-space and energy half-bandwidth, indicated by $D$. (c) Temperature exponents for scattering rate and resistivity with Hubbard repulsion $U$, sketched in (b), comparing 3D Dirac/Weyl semimetals against conventional Fermi liquids.}
\label{fig:cones_and_dos}
\end{figure}

The low-energy spectrum of Dirac and Weyl nodes in 3D cannot be gapped out by any symmetry-preserving single-particle perturbation. On the contrary, many-body effects, when dominant, lead to the breakdown of this protection and open a gap  \cite{Morimoto2016,Braguta2016}. We focus on the most fundamental and, at the same time, simplest case of electron-electron interaction: a local intra-orbital Hubbard repulsion $U$.
The energy-momentum dispersion and the density of states (DOS) of 3D Dirac semimetals are sketched in Fig.~\ref{fig:cones_and_dos}.
The half-bandwidth $D$ corresponds to an ultraviolet cutoff $\Lambda$ on the momenta and sets the energy scale from which the physics becomes non-perturbative, eventually leading to a Mott transition of the Dirac cone.   
In this Letter, we investigate the temperature dependence of the scattering rate as well as of the bulk diagonal electrical and thermal resistivities, from small to larger values of the dimensionless $U/D$ parameter, by means of analytical and non-perturbative numerical calculations.

In the weak-coupling limit, we explicitly test the applicability to 3D Dirac semimetals of the Landau Fermi-liquid theory, which is based on a one-to-one correspondence between a system of interacting electrons and a gas of asymptotically free fermions. 
In Fermi liquids (FL), quasiparticle excitations are well defined if their characteristic energy is larger than their inverse lifetime, proportional to the scattering rate.  If only electron-electron interaction is present the resistivity vanishes at zero temperature and grows as $T^2$. In 3D Dirac/Weyl semimetals, the DOS goes however quadratically to zero, approaching the Fermi level. 
This affects both the lifetime and the effective number of carriers available for transport leading to a characteristic and unexpected $T^6$-behavior for the resistivity $\rho$.
Interestingly, even away from the neutrality point ($E\!=\!0$), the power-law exponents remain fairly large (4 to 5) in a narrow though resolvable window of dopings. This observation has relevant implications for transport properties of Dirac/Weyl materials, such as WP$_2$, MoP, Cd$_3$As$_2$, as well as Hg$_{1-x}$Cd$_x$Te.

\noindent
{\it Weak-coupling result} -- 
In the subspace of spin and orbital degrees of freedom, denoted by the Pauli matrices $\vec{S}$ and $\vec{\tau}$ respectively, the three-dimensional non-interacting Hamiltonian considered in Fig.~\ref{fig:cones_and_dos} reads
\begin{eqnarray}
H_0= \hbar v   S_z \otimes (\vec{k} \cdot \vec{\tau}),
\label{H0}
\end{eqnarray}
where the momenta $\vec{k}$ live inside the sphere of radius $\Lambda$ and $v$ represents the velocity, i.e. the slope of the linear dispersion. 
The Hubbard $U$ acts when the same orbital is occupied by two electrons with opposite spins, i.e. here we do not consider inter-orbital two-body terms.

In the weak coupling regime ($U \! \ll  \! D$), perturbation theory holds and the scattering rate $\Gamma(\omega) = - \operatorname{Im} \Sigma(\omega)$ can be expressed as 
\begin{multline}  \label{eq:gamma_energy_integral}
    \Gamma(\omega) \propto \frac{\pi V^3 U^2}{( \hbar v)^9}  \int d \varepsilon_1 d\varepsilon_2 d\varepsilon_3 \delta(\omega +\varepsilon_1 - \varepsilon_2 - \varepsilon_3) \\\quad \times(f_1(1-f_2)(1-f_3) + (1-f_1)f_2 f_3) \, \varepsilon_1^2\, \varepsilon_2^2 \,\varepsilon_3^2,
\end{multline}
where $V$ is the volume of the unit cell, $f_i$ indicates the Fermi-Dirac distribution function of $\varepsilon_i$ and the cutoff has been set to infinity because of the weak-coupling condition. The quadratic behavior of the DOS enters explicitly in the integrand (last three factors) and cannot be considered constant any more, as in conventional metals.
An analytic evaluation of (\ref{eq:gamma_energy_integral}) yields \cite{suppl}
\begin{equation} \label{eq:analytic_gamma}
	\Gamma (\omega) \propto \frac{8 \pi V^3 U^2 T^8}{( \hbar v)^9} P_8(x)
\end{equation} 
where $x=\omega/T$ and $P_8(x)=\frac{x^8}{8!} + \frac{7 \pi^4 x^4 }{960}+\frac{31 \pi^6 x^2 }{1008} + \frac{3 \pi^8}{128}$, with $T$ implicitly including the Boltzmann constant $k_B$.
The first unexpected outcome of this simple calculation is therefore the much higher exponents entering the temperature and energy dependence of $\Gamma$, compared to the standard quadratic Fermi-liquid case \cite{coleman_introduction_2015}. 

Within the Kubo formalism we can compute the conductivity $\sigma$ making use of our analytical result (\ref{eq:analytic_gamma}) for $\Gamma(\omega)$ and relying on the bubble approximation \cite{vertexcorrections}.
Calculations can be further simplified by disregarding the effect of the real part of the self-energy, as done in Ref.~\cite{tabert_optical_2016} and rewriting the conductivity in a Drude-Boltzmann fashion.
This is achieved upon introducing the ``$f\omega^2$''-average of a quantity $\mathcal{Q}(\omega)$ 
\begin{equation}
\langle \mathcal{Q} \rangle_{f\omega^2} = 
\int \! d \omega  \left[ -\frac{\partial f}{\partial \omega} \right] \omega^2 \mathcal{Q}(\omega) \bigg/
\! \! \underbrace {\int \!  d \omega  \left[ -\frac{\partial f}{\partial \omega} \right]  \omega^2}_{\mathcal{N}},
\end{equation}
which leads to the definition of an effective number density $n_\text{eff}/m^*=\mathcal{N} /(6 \pi^2 \hbar^3 v)$
as well as
a scattering time
\begin{equation}
\label{tau}
\tau=\hbar \bigg\langle \frac{1}{2\Gamma(\omega)} + \frac{3 \Gamma(\omega)}{2 \omega^2}   \bigg\rangle_{f\omega^2}.
\end{equation}
Using these definitions, the conductivity assumes the simple Drude form:
\begin{equation}
\label{Drude}
\sigma =  \frac{ e^2 \tau n_{\text{eff}}  }{m^*}.
\end{equation}
This Drude-like formulation allows to disentangle the role of the DOS entering in the factor  $n_{\text{eff}}/m^*$ from that of the interaction, leading to a finite scattering time $\tau$ \cite{meff}.
In conventional metals, the temperature dependence of $\sigma$ stems entirely from $\tau$, as all other quantities in (\ref{Drude}) do not depend on $T$.
In contrast, the parabolic DOS of the 3D Dirac semimetal results in a quadratic behaviour of the effective number density: $n_\text{eff}/m^* = T^2 / (18 \hbar^3 v)$.
Using the polynomial expression of $\Gamma$ given in Eq. (\ref{eq:analytic_gamma}) and considering that  
at low temperature the contribution of the second term in Eq. (\ref{tau}) is irrelevant, we arrive at $\tau \propto T^{-8}$. The resistivity calculated from Eq.(\ref{Drude}), i.e. taking into account the temperature dependence of both $n_\text{eff}$ and of $\tau$, shows hence the characteristic $\rho \propto T^{6}$, shown in the table of Fig.~\ref{fig:cones_and_dos}.

\noindent
{\it Beyond weak-coupling} -- 
By releasing the constraint $U \!\ll\! D$ and replacing perturbative analytic approaches with numerical many-body methods, we can access the intermediate-to-strong-coupling region.
We do so using fully dressed dynamical mean field theory (DMFT) \cite{georges_dynamical_1996} Green's functions in the Kubo expression \cite{suppl}. As impurity solver, we use the iterated perturbation theory solver  \cite{georges_hubbard_1992,rozenberg_mott-hubbard_1994,kajueter_doped_1996,kajueter_new_1996}, which offers computational advantages maintaining a fair accuracy for our purposes \cite{suppl}.
\begin{figure}[th]
\centering
\includegraphics[width=\linewidth]{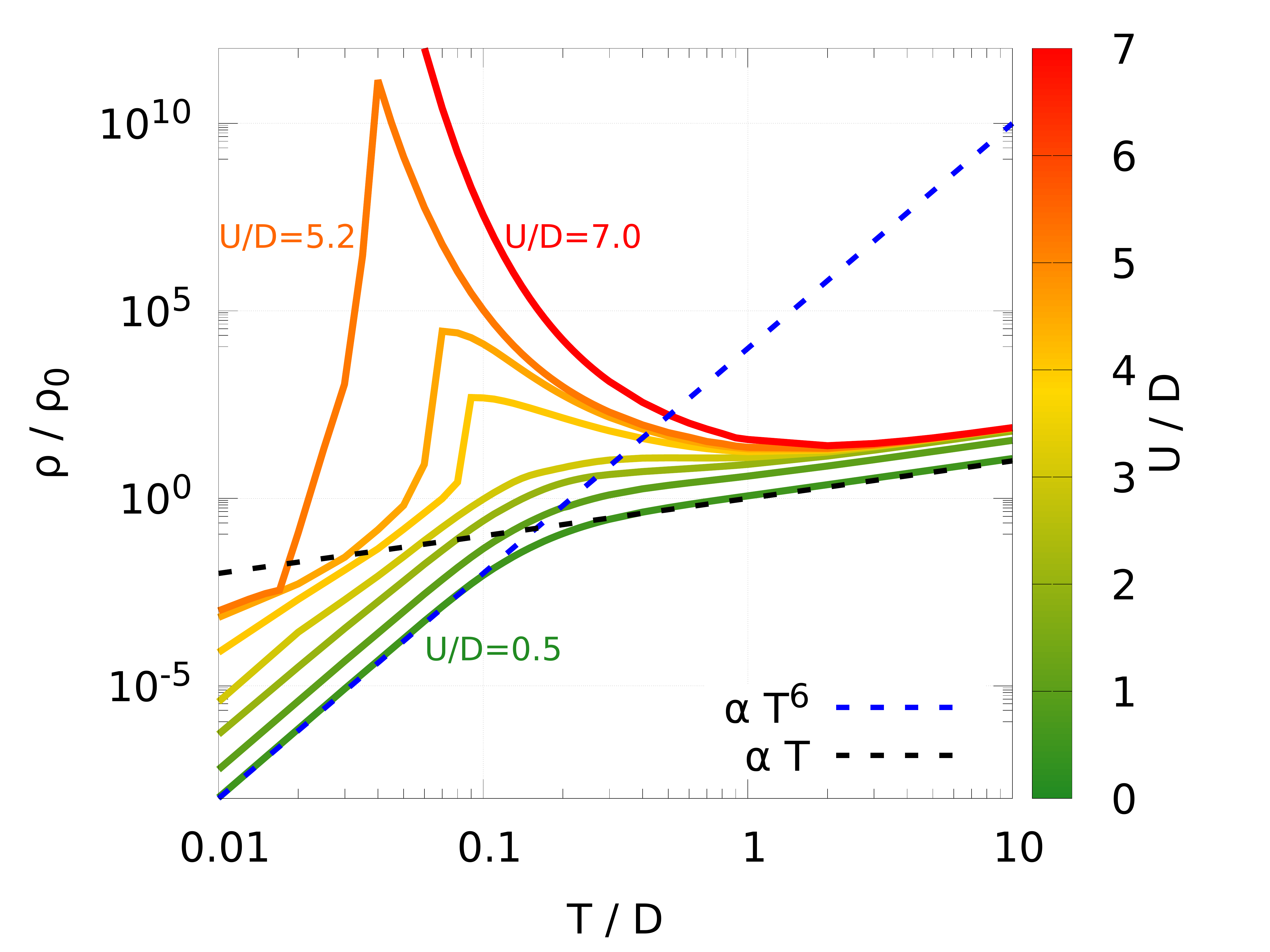}
\includegraphics[width=1\linewidth]{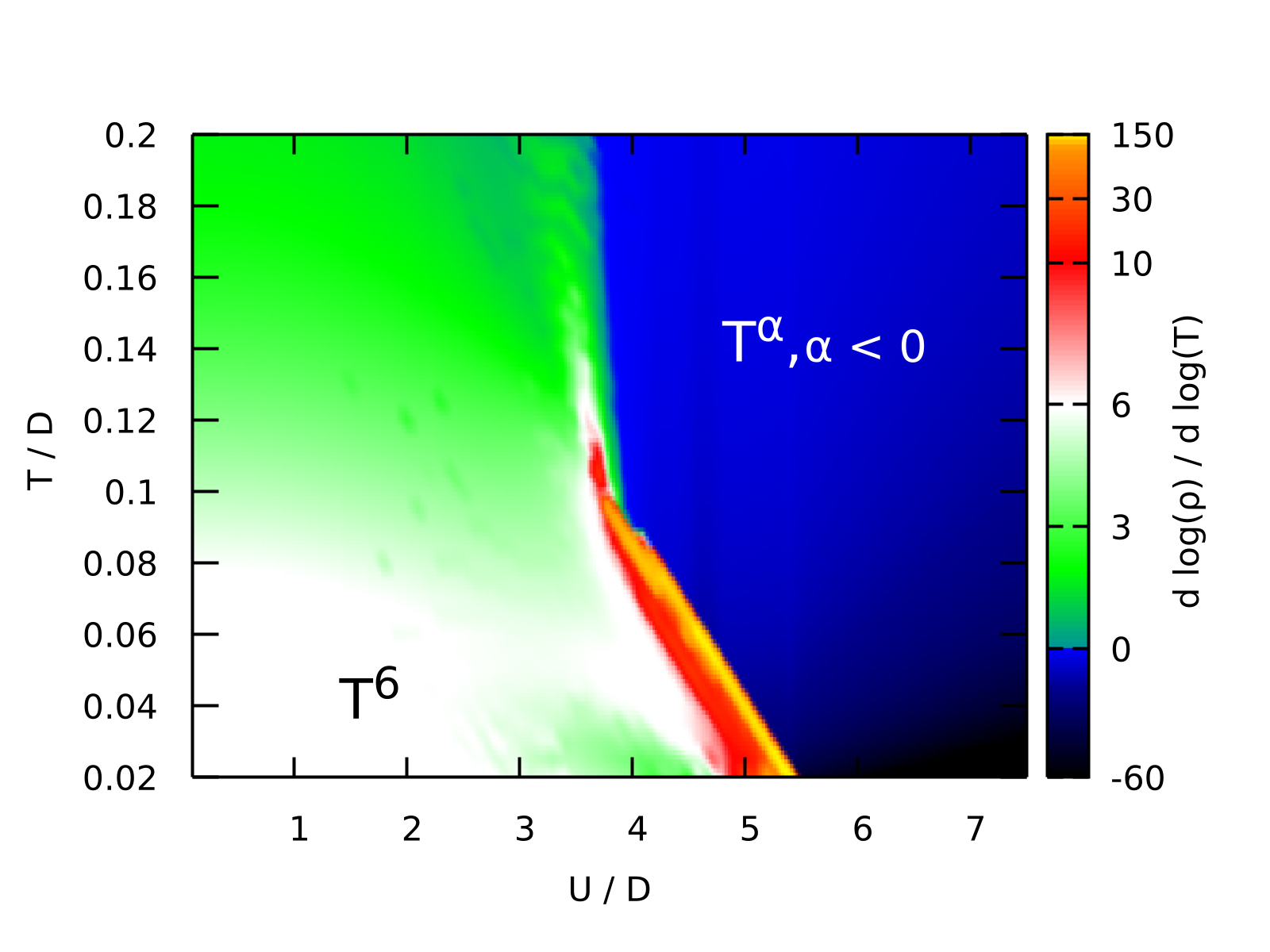}
\caption{Top panel: DMFT results for the resistivity $\rho$ of a 3D Dirac/Weyl semimetal at zero doping, as a function of the temperature $T$ for different values of the interaction strength $U/D$ (color bar), in units of $\rho_0 \! =\! 6 \pi^2 \hbar^3 v / (e^2 D)$.  The dashed lines indicate the asymptotic behavior for low temperature and weak coupling and for large temperature, respectively.
Bottom panel: DMFT phase diagram, where the color indicates $\frac{\partial \log \rho}{\partial \log T} $, i.e. the exponent of the $T$-dependence of $\rho$.}
\label{fig:rho_IPT}
\end{figure}
The resistivity $\rho$ as a function of $T$ for different values of $U$ and the corresponding exponent $\frac{\partial \log \rho}{\partial \log T}$ are shown in Fig.~\ref{fig:rho_IPT}. 
For small $U$, where DMFT nicely reproduces our analytic results (green curves and blue dashed line) as well as approaching the Mott transition located at $U_c\approx 5.5D$, we find that $\rho$ scales as $T^6$, in striking contrast with the $T^2$-resistivity of conventional metals.
This particular behavior is not limited to low temperatures but remains visible up to $T \approx 0.1 D $, as marked by the white region in the phase diagram in Fig.~\ref{fig:rho_IPT}.
The green region at low temperature around $U / D \approx 4$ reflects the presence of a low-energy kink in the $T^6$-resistivity behavior (see \cite{suppl}).
Upon further increasing $T$ at $U \!< \! U_c$ the resistivity displays a smooth crossover to a linear regime (green lines in Fig.~\ref{fig:rho_IPT} and dark green region in the phase diagram).
Let us mention, in passing, that bad-metal behavior characterized by $\rho \! \propto \! T$ has been reported in several other situations, from unconventional superconductors to theories of hydrodynamic transport \cite{lucas_resistivity_2017}, in which however the linear regime extends down to zero temperature. 

On the other side of the Mott transition instead, the typical exponential behavior of an insulating resistivity is recovered (red line in the upper panel of Fig.~\ref{fig:rho_IPT} and blue areas in the phase diagram).
Close to the Mott transition the exponent becomes very large, in analogy with the results for a semicircular DOS \cite{vucicevic_finite-temperature_2013}.
The parameter region where $\rho \propto T^6$ holds, is substantially larger than the corresponding $T^2$-one of conventional FL \cite{vucicevic_finite-temperature_2013}. Since this is  controlled by the Kondo scale of the emergent local moments, we can infer that the Kondo screening in Dirac and Weyl semimetals is strongly modified compared to standard metals.

\noindent
{\it Finite doping} -- 
After having found the $T^{-6}$ dependence of the electrical conductivity, we examine how far this persists upon doping the Dirac semimetal ($\mu \! \neq \! 0$).
Already a small doping adds contributions to the scattering rate with lower exponents, down to $T^2 \mu^6$. For $\omega \! =\! 0$, an analytic calculation yields \cite{suppl}
		\begin{equation} \label{eq:analytic_gamma_rate_doped}
		\Gamma (\omega\!=\!0) \propto T^8 Q_6(y),
		\end{equation} 
where now $y=\mu/T$ and $Q_6(y)=3 \pi^8  \!+\! 12 \pi^6  y^2 \!+\!20 \pi^4 y^4 \!+\! 8 \pi^2 y^6$.
When the temperature scale is substantially smaller than the chemical potential, i.e. $T \! \lessapprox\! 0.2\, \mu(T)$, the thermal broadening is negligible and the Dirac point is sufficiently away in energy, such that it plays no role for transport properties. In this case, the $T^2$ term in Eq.~(\ref{eq:analytic_gamma_rate_doped}) dominates and the scattering rate behaves similar to usual FL. When $T$ is increased beyond $0.2\,\mu$, the higher-order terms begin to contribute significantly. At $T \!>\! 0.6\, \mu$ (i.e. when the standard deviation of the derivative of the Fermi distribution ($\approx \! 1.8\, T$) is roughly equal to $\mu$) the $T^8$ term wins.	

\begin{figure}[th]
\centering
\includegraphics[width=1\linewidth]{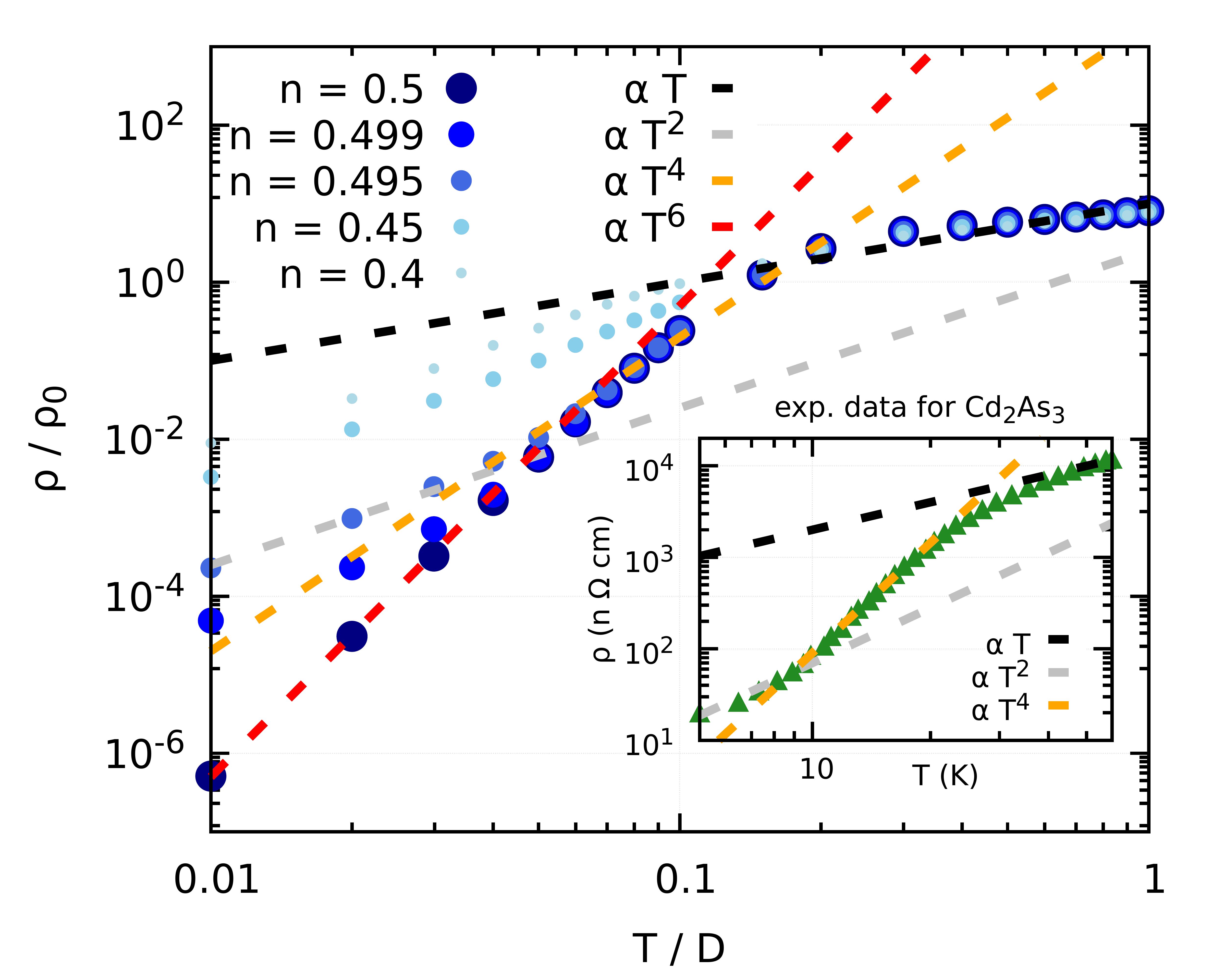}
\caption{DMFT results for the resistivity as a function of temperature of a Dirac semimetal for different fillings at interaction strength $U=2D$. The inset shows data for Cd$_3$As$_2$ taken from Ref.~\cite{liang_ultrahigh_2015}.}
\label{fig:resistivity_doped}
\end{figure}

This is confirmed by a DMFT calculation of $\rho$ away from the neutrality point, shown in Fig.~\ref{fig:resistivity_doped}.
Upon increasing the temperature at $n\!=\!0.5$ (dark blue circles), $\rho$ goes from $T^6$ directly to $T$.  
With a small but finite doping, the conventional $T^2$ Fermi-liquid behavior emerges at low temperatures, in agreement with the analytical result (see Fig.~\ref{fig:resistivity_doped}). At larger temperatures, however, before crossing over to the linear behavior, higher $T$-exponents in $\rho$ are clearly visible. The extension of this intermediate region depends on the doping level.
As we will discuss below, the possibility of realizing fast-growing (even though not necessarily a clean $T^6$) power-law resistivity can play a crucial role for the explanation of experiments on doped Dirac and Weyl semimetals.
For larger deviations from half-filling the resistivity goes instead directly from quadratic to linear and the influence of the Dirac fermiology in the intermediate region is suppressed. 

\noindent
{\it Other sources of scattering} --
The precise way in which $\Gamma$ depends on $\omega$ stems from the nature of the one- and two-body terms that we include in the model Hamiltonian. These can be, for instance, a random disorder potential, an electron-phonon interaction, as well as different parametrizations of the electron-electron repulsion, like the intra-orbital Hubbard $U$ considered here. 

Within the self-consistent Born approximation (SCBA), Im$\Sigma$ arising from disorder is proportional to the DOS -- hence it is quadratic in $\omega$ -- and to the variance of the disorder distribution. As a consequence $\tau \! \propto \! T^2$.
Interband transitions can be safely neglected.
This way the temperature in the expression for $\sigma$ drops out, yielding a residual resistivity at low $T$.
The case of (a weak) electron-phonon interaction is similar, the only difference being an explicit (linear) temperature dependence appearing in the variance of the distribution of local displacements associated to classical phonons above the Debye temperature. 
This leads to $\tau \! \propto \! T^3$ and, upon plugging this into (\ref{Drude}), 
to $\sigma \! \propto \! 1/T$. 
A third situation which can be easily recasted in this simplified Drude-Boltzmann-like framework is a long-range Coulomb interaction at weak coupling strength: in this case, $\Gamma(\omega)$ is given by $\text{max}(\omega, T)$ \cite{burkov_topological_2011,hosur_charge_2012}. Interband contributions matter and one obtains $\tau \! \sim \! 1/T$ leading to $\sigma \! \propto \! T$ \cite{suppl}.
Having re-obtained these results make the comparison of the various scattering mechanisms in three spatial dimensions easy. Vertex corrections may introduce modifications as it has been shown, for example, for Dirac fermions in 2D, where transport and quasiparticle scattering times get different $T$-dependences \cite{lara_bft_2009}.

\noindent
{\it Specific heat and thermal transport} -- 
The dispersion of Dirac/Weyl semimetals affects also other important transport and thermodynamic coefficients. One relevant example is the temperature behavior of the speficic heat. 
In contrast to the usual renormalized linear behavior of a conventional FL ($c_\text{V} \propto T/Z$, with $Z$ being the quasiparticle weight) for the undoped Dirac semimetals it depends on the cube of the temperature ($c_\text{V} \propto T^3/Z^3$) \cite{lai_weylkondo_2018}. 
Similar as for $\rho$, doping away from the degeneracy point introduces a conventional (linear) behaviour at low temperatures:
$c_\text{V}(T) \propto \frac{7 \pi^4 T^3}{5 Z^3}(1 + a y^2)$
with $y=\mu/T$ and $a=\frac{5 Z^2}{7 \pi^2}$.
The crossover between FL-like and Dirac-like behavior happens at $T\approx 0.27 Z \mu$, 
i.e. approximately the same temperature as for the scattering rate. 
We calculated the specific heat within DMFT using the approach outlined in Ref.~\cite{rohringer_impact_2016}. The results confirm the trend given by the analytic expression \cite{suppl}. 

Similarly, our DMFT calculations of the thermal conductivity yield results in sharp contrast to conventional FL: we find $\kappa \propto T^{-5}$.  
In analogy to $1/\sigma$ shown in Fig.~\ref{fig:rho_IPT}, the thermal resistivity $1/\kappa$ displays a crossover at $T \sim 0.1D$, from $T^5$ to $T^2$.   
We can then conclude that the $\kappa/\sigma$ ratio is linear in $T$ at low temperatures, which represents a result compatible with a Wiedemann-Franz-type of relation. 
Let us recall that this conclusion holds roughly in the white region of the phase diagram of Fig.~\ref{fig:rho_IPT} and is reached within the bubble approximation for the conductivities \cite{suppl}.
\noindent
{\it Relation to materials} -- 
First of all it is important to stress that, even though we have so far been explicitly referring to 3D Dirac semimetals, our results apply also to the case of Weyl nodes, since these are characterized by the same quadratic DOS. 
In the literature, high exponents for the resistivity have been reported in some Dirac and Weyl semimetals, e.g. in the type-II Weyl system WP$_2$ \cite{kumar_extremely_2017}, in the multifold-fermion system MoP \cite{kumar_extremely_2019} and in the 3D Dirac semimetal Cd$_3$As$_2$ \cite{liang_ultrahigh_2015}. Data for the latter is shown in the inset to Fig.~\ref{fig:resistivity_doped} (see \cite{suppl} for the other compounds WP$_2$ and MoP). As discussed above for the finite doping case, already a small distance between the Fermi energy and the Dirac point leads to the emergence of a conventional $T^2$-behaviour for small $T$. 
Our results remain however applicable to the intermediate temperature-region shown in Fig.~\ref{fig:resistivity_doped},
providing an interpretation for the observed $T^4$ behaviors of $\rho$ without any {\it ad-hoc} assumption.

Hg$_{1-x}$Cd$_x$Te has been reported as a realization of the so-called Kane-semimetal \cite{orlita_observation_2014}, i.e. a 3D zero-gap Dirac system. Provided that the influence of the structural thermal expansion remains small, we expect the appearence of $T^6$ terms in the longitudinal resistivity. The situation at the critical doping, where the gap closes, is however disturbed by the presence of an additional flat band. 
The Weyl semimetal realized in compressively strained HgTe might offer a prosiming alternative, as it avoids the disturbance of the heavy-hole band and requires no fine-tuning of the doping \cite{mahler_interplay_2019}.

\noindent
{\it Conclusions} -- 
In three-dimensional Dirac and Weyl semimetals, the temperature power law of the scattering rate originating from short-range electron-electron interaction gains as many as \emph{six} powers, compared to conventional FL.    
As a consequence, electrical and thermal resistivities are altered dramatically and go up \emph{four} powers of $T$: in Dirac/Weyl systems we indeed find $\rho \! \propto \! T^6$ and $\kappa^{-1} \! \propto \! T^5$.
These conclusions hold not only for weak strengths of the Hubbard repulsion, where they can also be derived analytically, but also in a substantial region of the $T$-$U$ phase diagram, essentially up to the point at which the semimetal breaks down and is turned into a Mott insulator.   
These particular temperature exponents lead to an unusual situation: they describe a strongly suppressed contribution to transport from electron-electron repulsion at low temperatures but they rapidly prevail upon increasing $T$, in the intermediate-to-high regime. 
Interestingly, if we move away from the Dirac/Weyl point, the temperature exponent for the electrical resistivity is only slightly altered, shifting to $\rho \propto T^{4-5}$, offering a natural and simple explanation to the so far elusive origin of the high exponents of bulk diagonal transport coefficients measured in several Dirac and Weyl materials.   

\noindent
\begin{acknowledgements}

{\it Acknowledgments} -- 	
We thank D. Di Sante for useful conversations.  
N.W., B.T. and G.S. are supported by the SFB 1170 Tocotronics,
Funded by the Deutsche Forschungsgemeinschaft (DFG, German Research Foundation) – Project-ID 258499086 – and acknowledge financial support from the DFG through the Würzburg-Dresden Cluster of Excellence on Complexity and Topology in Quantum Matter–{\it ct.qmat} (EXC 2147, project-id 390858490). 
We gratefully acknowledge the Gauss Center for Supercomputing e.V. (www.gauss-center.eu) for funding this project by providing computing time on the GCS Supercomputer SuperMUC at Leibniz Supercomputing Center (www.lrz.de). Computing time at HLRN (Berlin and Göttingen) is acknowledged.
A.T. acknowledges financial support from the Austrian Science Fund (FWF), through the project I 2794-N35.
\end{acknowledgements}
\bibliography{references.bib}

\begin{thebibliography}{43}
\bibitem [{ver()}]{vertexcorrections}%
  \BibitemOpen
  \bibinfo {note} {Restricting to the bubble diagram does not represent a too
  severe approximation in our case, as we are going to compute the
  conductivities of these 3D semimetals within dynamical mean-field theory and
  with an intra-orbital-only Hubbard interaction, for which the vanishing of
  vertex corrections in ladder diagrams can be explicitly demonstrated
  \cite{suppl}. Adding inter-orbital two-body terms would instead generate
  vertex corrections to the current-current response, similarly to what occurs
  for disorder or electron-phonon interaction}\BibitemShut {NoStop}%
\bibitem [{mef()}]{meff}%
  \BibitemOpen
  \bibinfo {note} {Note that $m^*$ is equal to $m$ if one makes exactly the
  same approximations as in \cite{tabert_optical_2016}, however the formulas
  given here are more general since one can in principle include the effects of
  the real part of the self-energy via $m^*$.}\BibitemShut {Stop}%
\end{thebibliography}

\end{document}


	\title{Supplemental Material\\ \vspace*{0.3cm} Resistivity Exponents in 3D-Dirac Semimetals From Electron-Electron Interaction}
	
\author{Niklas Wagner}
	\affiliation{Institut f\"ur Theoretische Physik und Astrophysik, Universit\"at W\"urzburg, 97074 W\"urzburg, Germany}
	\author{Sergio Ciuchi}%
	\affiliation{Dipartimento di Scienze Fisiche e Chimiche, Universit`a dell’Aquila, and Istituto dei Sistemi Complessi, CNR, Coppito-L’Aquila, Italy}
	\author{Alessandro Toschi}
	\affiliation{Institute of Solid State Physics, TU Wien, 1040 Vienna, Austria}
	\author{Björn Trauzettel}
	\affiliation{Institut f\"ur Theoretische Physik und Astrophysik and W\"urzburg-Dresden Cluster of Excellence ct.qmat, Universit\"at W\"urzburg, 97074 W\"urzburg, Germany}
	\author{Giorgio Sangiovanni}
	\affiliation{Institut f\"ur Theoretische Physik und Astrophysik and W\"urzburg-Dresden Cluster of Excellence ct.qmat, Universit\"at W\"urzburg, 97074 W\"urzburg, Germany}

	\begin{abstract}

	\end{abstract}

	\maketitle
	
\section{Numerical calculations}
    \subsection{Iterated Perturbation Theory}
        Iterated Perturbation Theory(IPT) \cite{georges_hubbard_1992,rozenberg_mott-hubbard_1994,kajueter_doped_1996} is a simple implementation of dynamical mean-field theory(DMFT) \cite{georges_dynamical_1996}. IPT has the advantage that we can carry out the calculations on the real axis and therefore we don't have to rely on analytic continuation. We use a self-consistency scheme which consists of three steps:\\
        First the local Greens function is computed ($\mu=0$ at half-filling and we set $\hbar=1$):
        \begin{equation} \label{eq:gloc}
            G_{\textbf{loc}}(\omega) = \frac{1}{N} \sum_{\bm{k}} [\omega -H(\bm{k}) +\mu -\Sigma(\omega)]^{-1}.
        \end{equation}
        Then
        \begin{equation}
            G_0=(G_{\text{loc}}^{-1} + \Sigma(\omega))^{-1}
        \end{equation}
        is computed. Finally, a new self-energy is computed 
        \begin{equation} \label{eq:sigma}
            \Sigma(\omega) = -\pi U^2 \int dx dy \rho(\omega -x) \rho(y) \rho(y-x) (f(\omega-x) f(y) f(x-y) + f(x-\omega) f(-y)f(y-x)).
        \end{equation}
        and the loop is closed by computing a new $G_{\textbf{loc}}(\omega)$ using eq.\eqref{eq:gloc}. Here $\rho(x)= -\frac{1}{\pi} \operatorname{Im} G_0(x)$ and $f$ is the Fermi function.\\
        In the case of the Dirac semimetal model  defined in the main text, the Greens function is proportional to the identity matrix and Eq.\eqref{eq:gloc} can be rewritten as an integral over energy which can be solved analytically. In that case
        \begin{equation}
            G_{\textbf{loc}}(z) = 3 (-z^2 \log((z - D) / (z + D)) - 2 z D) / 2
        \end{equation}
        with $z= \omega -\Sigma(\omega) +i \delta$ and cutoff $D$.\\
        For finite doping a modified version of IPT has been proposed \cite{kajueter_new_1996,kajueter_doped_1996}. In this case the self-energy ansatz is the following:
        \begin{equation}
         \Sigma_{\text{doped}}(\omega)=   U n +  \frac{\frac{n(1-n)}{n_0(1-n_0)} \Sigma(\omega)}{1- \frac{(1-n)U-\mu+\tilde{\mu}}{n_0(1-n_0) U^2} \Sigma(\omega)}
        \end{equation}
        where $\Sigma$ is calculated using eq.\eqref{eq:sigma}. $\mu$ is the chemical potential and $n$ the filling. $n_0$ is the filling associated with 
        \begin{equation}
            G_0=(G_{\text{loc}}^{-1} + \Sigma(\omega)-\mu + \tilde{\mu})^{-1}
        \end{equation}
        which replaces $G_0$ in the self-consistency loop above. $\tilde{\mu}$ is a free parameter which can be fixed by requiring that $n_0=n$ \cite{potthoff_interpolating_1997}.\\
        Because the self-energy is approximated using the second order diagram, IPT becomes correct in the weak-coupling limit. For very strong coupling, i.e. in the atomic limit, the results IPT produces become exact again \cite{rozenberg_mott-hubbard_1994}. For intermediate values of the interaction strength IPT is known to give qualitatively correct results with some quantitative deviations of the critical coupling for the Mott transition with respect to exact solver. 
   \subsection{Electrical conductivity}
\begin{figure}[h]
            \centering
            \includegraphics[width=0.5\linewidth]{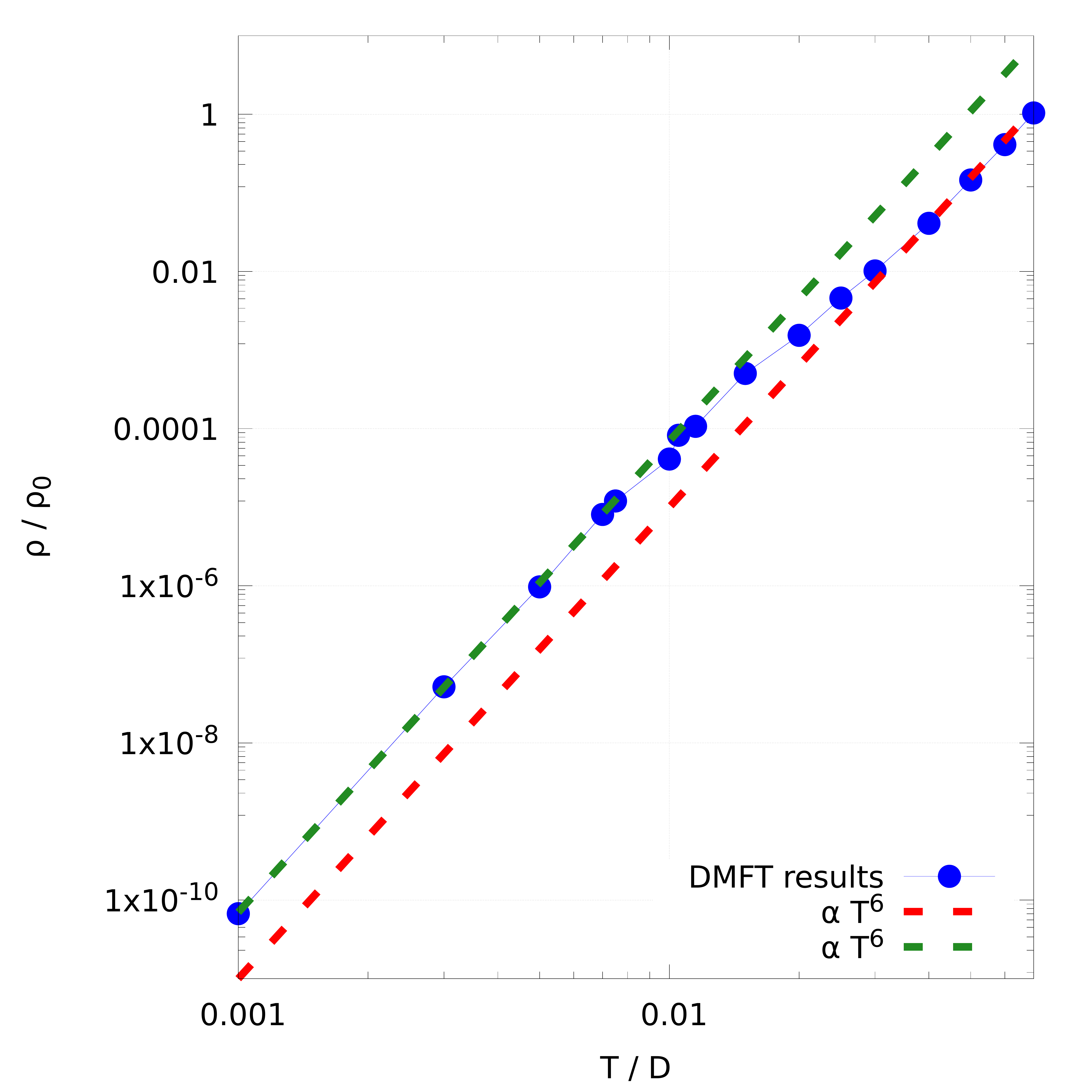}
            \caption{DMFT results for the resistivity at $U / D = 4$. Here $\rho_0=\frac{6\pi^2 \hbar^2 v}{e^2 D }$ with half-bandwidth $D$. The dashed lines show the asymptotic behaviour before and after the kink.}
            \label{fig:kink}
        \end{figure}

       Using the self-energy obtained with DMFT the conductivity is computed  using a Kubo formula approach (see also sec. \ref{sec:vtx})\cite{tabert_optical_2016,ashby_chiral_2014}:
    		\begin{eqnarray} \label{eq:kubo}
    		\sigma=\frac{e^2}{6\pi^2 \hbar^2 v} \int d \omega \left(- \frac{\partial f}{\partial \omega}\right) \int d \epsilon \, \epsilon^2 \left(A_+^2 + 2 A_+ A_-\right)
    		\end{eqnarray}
    		
    where $A_\pm=-\frac{1}{\pi} \operatorname{Im} \frac{1}{\omega \mp \epsilon -\Sigma(\omega)} $, $v$ is the velocity parameter and self-energy $\Sigma(\omega)$. The first term in the integrand describes intraband-scattering and the second term interband-scattering. The results obtained with this approach are shown in the main text in Fig 2. Note that for very small temperatures the results in the Mott phase are acquired by fitting and extrapolation.\\
    The green region around $U / D \approx 4$ in the phase diagram (Fig. 3 in the main text) reflects  a low-energy kink in the resistivity (see Fig. \ref{fig:kink}). Indeed, above $T /D \approx 0.04$ and below $T /D \approx 0.01$ the resistivity goes as $T^6$ but with different prefactors. Such kinks have been previously reported as a general consequence of strong electronic correlations in the proximity of a Mott-Hubbard transition, see Refs.~\cite{byczuk_kinks_2007,toschi_kinks_2009,PhysRevLett.110.246402}. The kinks described there however, affect the conventional Fermi-liquid behavior of the spectral and transport properties.
    \subsection{Thermal conductivity}	
\begin{figure}[h]
            \centering
            \includegraphics[width=0.7\linewidth]{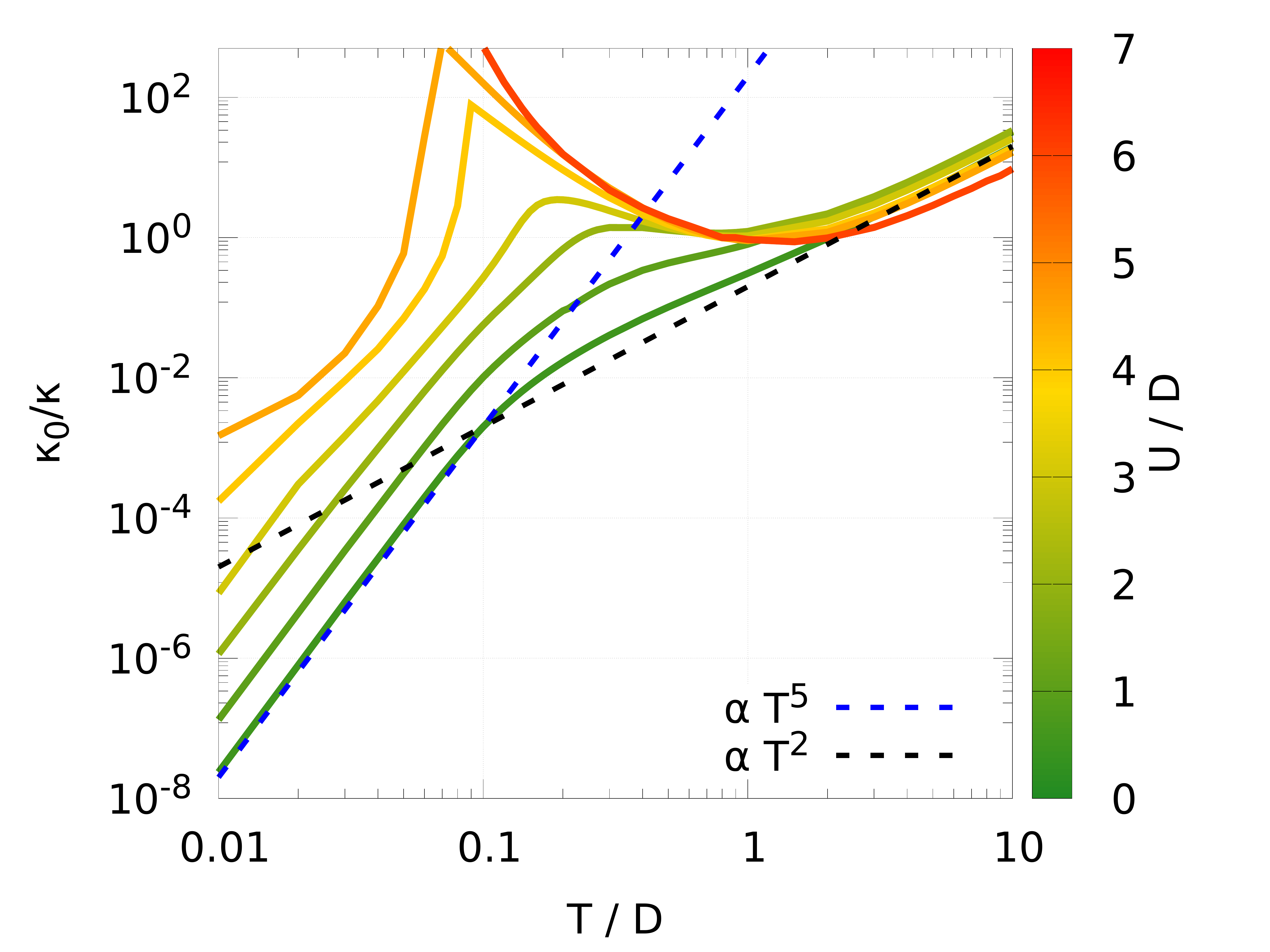}
            \caption{DMFT results for the thermal resistivity $\kappa^{-1}$ for different $U$ at half-filling. Here $\kappa_0= \frac{k_B D^2}{6 \pi \hbar^2 v}$ with the Boltzmann constant $k_B$ and half-bandwidth $D$. The crossover from $T^5$ to quadratic behavior is highlighted by the dashed guide-to-the-eye fits. The results shown in the Mott phase for small T are obtained using a fit of the DMFT results.}
            \label{fig:therm_res}
        \end{figure}

        We compute the thermal conductivity following Ref.~\onlinecite{oudovenko_thermoelectric_2002}, while taking inter- and intraband contributions into account:
         	\begin{eqnarray} \label{eq:thermal}
            	\kappa= \frac{k_B}{6 \pi \hbar^2 v} \int d \omega \left(\frac{\omega}{T}\right)^2\left(- T \frac{\partial f}{\partial \omega}\right) \int d \epsilon \, \epsilon^2 \left(A_+^2 + 2 A_+ A_-\right)
        	\end{eqnarray}
        with $A_\pm=-\frac{1}{\pi} \operatorname{Im} \frac{1}{\omega \mp \epsilon -\Sigma(\omega)} $, self-energy $\Sigma(\omega)$  and Boltzmann constant $k_B$. Note that we include a factor of $k_B$ in $T$.\\
        Inserting the DMFT results for the self-energy in the above formula leads to the results displayed in Fig. \ref{fig:therm_res} and discussed in the main text.
            
\section{Analytic expression for the imaginary part of the self-energy at weak coupling}
            For a Dirac semimetal the derivation of the scattering rate (i.e. the imaginary part of the self-energy) has to be changed compared to the FL case \cite{coleman_introduction_2015} because we have to take the parabolic DOS into account.
        \subsection{Half-filling}

            Based on the second-order diagram and using the Dirac semimetal density of states  $N(\omega)=\frac{1}{( \hbar v)^3}\omega^2$ (which has dimensions $1/[\text{volume}\cdot \text{energy}]$)  the imaginary part of the local self-energy can be expressed as:
            \begin{equation} \label{eq:gamma_energy_integral}
            \Sigma^{''}(\omega) \propto -\frac{\pi V^3}{( \hbar v)^9} U^2 \int d \varepsilon_1 d\varepsilon_2 d\varepsilon_3 \delta(\omega +\varepsilon_1 - \varepsilon_2 - \varepsilon_3) (n_1(1-n_2)(1-n_3) + (1-n_1) n_2 n_3) \varepsilon_1^2 \varepsilon_2^2 \varepsilon_3^2
            \end{equation}
            where $n_i=f(\epsilon_i)$ is the Fermi function and $V$ is the volume of the unit cell. The proportionally symbol $\propto$ does not include any dimensional factor. 
            This can be rewritten as:
            \begin{equation}
            \Sigma^{''}(\omega) \propto-\frac{\pi V^3}{( \hbar v)^9} U^2 \left( \xi(\omega)+\xi(-\omega)\right) 
            \end{equation}
            with
            \begin{equation} \label{eq:energy_integral}
            \begin{aligned}
            \xi(\omega)&= \int d \varepsilon_1 d \varepsilon_2 d \varepsilon_3 \delta(\omega -(\varepsilon_1 + \varepsilon_2 + \varepsilon_3))n_1 n_2 n_3 \varepsilon_1^2 \varepsilon_2^2 \varepsilon_3^2\\&=
             \int \frac{d \alpha}{2 \pi} d \varepsilon_1 d\varepsilon_2 d\varepsilon_3 e^{i \alpha(\omega -(\varepsilon_1 + \varepsilon_2 + \varepsilon_3))} n_1 n_2 n_3 \varepsilon_1^2 \varepsilon_2^2 \varepsilon_3^2\\&=
             \int \frac{d \alpha}{2 \pi} e^{i \alpha \omega} A(\alpha)^3.
            \end{aligned}
            \end{equation}
            Here $A(\alpha)$ has been introduced which is given by 
            \begin{equation}
            \begin{aligned}
            A(\alpha)&= \int d \varepsilon e^{-i\tilde{\alpha} \varepsilon} f(\varepsilon) \varepsilon^2
            \end{aligned}
            \end{equation}
            with $\tilde{\alpha}=\alpha+ i \delta $.\\
            For $\alpha\geq0$ we can close the integration path in the lower complex half-plane:
            \begin{equation}
            \begin{aligned}
            A(\alpha)&=-2 \pi i T \sum_{m=0}^{\infty} e^{-\tilde{\alpha} \pi T (2m+1)}(\pi T (2m+1))^2\\&=-
            \frac{i \pi^3 T^3}{2} \frac{(3+ \cosh(2 \pi T\tilde{\alpha}))}{\sinh(\pi T\tilde{\alpha})^3}.
            \end{aligned}
            \end{equation}
            For $\alpha<0$ the same result is obtained when closing the path in the upper half-plane.\\
            $A(\alpha)$ has poles at $\alpha +i \delta= \frac{i n}{T}$ ($n \in N$) and vanishes at infinity. Therefore we we can evaluate eq.\eqref{eq:energy_integral} by closing the contour in the upper half-plane for $\omega>0$. Expanding $A(\alpha)^3$ in the vicinity of the $n$-th pole yields:
            \begin{equation}
            A_n(\alpha)^3= 8 i (-1)^n \left( \frac{1}{(\tilde{\alpha}- \frac{i n}{T})^9} + \frac{7 \pi^4 T^4}{40 (\tilde{\alpha}- \frac{i n}{T})^5} -\frac{31 \pi^6 T^6}{504 (\tilde{\alpha}- \frac{i n}{T})^3} + \frac{3 \pi^8 T^8}{128 (\tilde{\alpha}- \frac{i n}{T})} + \mathcal{O}(\tilde{\alpha}- \frac{i n}{T})\right).
            \end{equation}
            Now the integral over the poles can be carried out:
            \begin{equation}
            \begin{aligned}
            \xi(\omega)&= \int \frac{d \alpha}{2 \pi} e^{i \alpha \omega} A(\alpha)^3 = \oint \frac{d z}{2 \pi} e^{i z \omega} A(z)^3\\&=- 8 \sum_{n=1}^{\infty} (-1)^n e^{-n \omega/T} \left( \frac{\omega^8}{8!} + \frac{7 \pi^4 \omega^4 T^4}{40\cdot 4!} +\frac{31 \pi^6 \omega^2 T^6}{504 \cdot 2!} + \frac{3 \pi^8 T^8}{128} \right)\\&=
            + 8\frac{1}{1+ e^{\omega/T}} \left( \frac{\omega^8}{8!} + \frac{7 \pi^4 \omega^4 T^4}{40\cdot 4!} +\frac{31 \pi^6 \omega^2 T^6}{504 \cdot 2!} + \frac{3 \pi^8 T^8}{128} \right).
            \end{aligned}
            \end{equation}
            For $\omega<0$ we get a minus for closing in the lower half-plane but we also have to start the sum at $n=0$ because the poles are located at $z= \frac{i n}{T} - i \delta$. Thus in this case we have to take the pole at $n=0$ into account which leads to the same result as for $\omega >0$.\\
            Finally:
            \begin{equation} \label{eq:imsigma}
            \Sigma^{''}(\omega) \propto -\frac{\pi V^3}{( \hbar v)^9} U^2 \left( \xi(\omega)+\xi(-\omega)\right)  = -\frac{8 \pi V^3}{( \hbar v)^9} U^2 \left( \frac{\omega^8}{8!} + \frac{7 \pi^4 \omega^4 T^4}{40\cdot 4!} +\frac{31 \pi^6 \omega^2 T^6}{504 \cdot 2!} + \frac{3 \pi^8 T^8}{128} \right) .
            \end{equation}
    
    \subsection{Finite doping}
        For finite doping, 
        the DOS in Eq. \eqref{eq:gamma_energy_integral} has to be changed to $\varepsilon_i \rightarrow \varepsilon_i - \mu$.  Then the shift $\varepsilon_1\rightarrow- \varepsilon_1$ leads to
        \begin{equation}
        \begin{aligned}
        \Sigma^{''}(\omega)&\propto -\frac{\pi V^3}{( \hbar v)^9} U^2\int d \varepsilon_1 d\varepsilon_2 d\varepsilon_3 \delta(\omega -(\varepsilon_1 + \varepsilon_2 + \varepsilon_3)) ((1-n_1)(1-n_2)(1-n_3) +n_1 n_2  n_3) (\varepsilon_1+\mu)^2(\varepsilon_2-\mu)^2 (\varepsilon_3-\mu)^2\\&= -\frac{\pi V^3}{( \hbar v)^9} U^2 \big(\int d \varepsilon_1 d\varepsilon_2 d\varepsilon_3 \delta(\omega -(\varepsilon_1 + \varepsilon_2 + \varepsilon_3)) (n_1 n_2 n_3) (\varepsilon_1+\mu)^2(\varepsilon_2-\mu)^2 (\varepsilon_3-\mu)^2\\& \quad +\int d \varepsilon_1 d\varepsilon_2 d\varepsilon_3 \delta(\omega +(\varepsilon_1 + \varepsilon_2 + \varepsilon_3)) ( n_1 n_2 n_3) (\varepsilon_1-\mu)^2(\varepsilon_2+\mu)^2 (\varepsilon_3+\mu)^2 \big)\\&
        = -\frac{\pi V^3}{( \hbar v)^9} U^2 \left(\int  \frac{d \alpha}{2 \pi} e^{i \alpha \omega}A(\alpha,\mu)A(\alpha,-\mu)^2 +\int  \frac{d \alpha}{2 \pi} e^{-i \alpha \omega}A(\alpha,-\mu)A(\alpha,\mu)^2 \right)
        \end{aligned}
        \end{equation}
        with 
        \begin{equation}
        \begin{aligned}
        A(\alpha,\mu)&= \int d \varepsilon e^{- i\tilde{\alpha} \varepsilon} f(\varepsilon) (\varepsilon+\mu)^2\\&=
        - 2 \pi i T \sum_{n=0}^{\infty} e^{-\alpha \pi T (2n+1)}(i\pi T (2n+1)+ \mu)^2 .
        \end{aligned}
        \end{equation}
        As for the half-filling case the poles of this expression are located at $\tilde{\alpha} = \frac{i n}{T}$ which allows to calculate the $\alpha$ integral in the same way. This leads to
        \begin{equation}
            \begin{aligned}
            		\Sigma^{''}(\omega)&\propto -\frac{\pi V^3}{( \hbar v)^9} U^2 \Big[\frac{1}{16}\left(3 (\pi T)^8 + 12 (\pi T)^6 \mu^2 +20 (\pi T)^4 \mu^4 + 8 (\pi T)^2\mu^6\right)\\&
            		\quad+\omega \left(\frac{19}{210} (\pi T)^6 \mu +\frac{2}{3}(\pi T)^4 \mu^3 +\frac{1}{3} (\pi T)^2 \mu^5\right)+\omega^2\left(\frac{31}{126} (\pi T)^6  +\frac{169}{180}(\pi T)^4\mu^2+\frac{4}{3}(\pi T)^2 \mu^4 +\frac{1}{2}\mu^6\right)\\&
            		\quad +\omega^3 \left(\frac{7}{90} (\pi T)^4 \mu +\frac{2}{3} (\pi T)^2\mu^3 +\frac{1}{3}\mu^5\right)+\omega^4\left(\frac{7}{120}(\pi T)^4 +\frac{7}{36}(\pi T)^2\mu^2+\frac{1}{12}\mu^4\right)\\&
            		\quad -\omega^5 \frac{1}{90}(\pi T)^2\mu +\omega^6 \frac{1}{180}\mu^2+\omega^7\frac{1}{630}\mu+\omega^8 \frac{1}{5040}\Big].
            \end{aligned}
        \end{equation}

\section{Analytic results for the conductivity }
       Following Ref. \cite{tabert_optical_2016} the two terms appearing in  Eq. \ref{eq:kubo} yield
    \begin{eqnarray}
    \sigma^{intra}&=& \frac{e^2}{6\pi^2 \hbar^2 v}  \int d \omega \left(- \frac{\partial f}{\partial \omega}\right) \frac{(\omega-\Delta(\omega))^2+\Gamma^2(\omega)}{2\Gamma(\omega)}\label{eq:intra}\\
    \sigma^{inter}&=& \frac{e^2}{6\pi^2 \hbar^2 v}  \int d \omega \left(- \frac{\partial f}{\partial \omega}\right) \Gamma(\omega)\label{eq:inter}
    \end{eqnarray}
    where
    $\Gamma(\omega)=-\operatorname{Im}\Sigma(\omega)$ and $\Delta(\omega)=\operatorname{Re}\Sigma(\omega)$
    We sum the two terms and recast the result by dividing and multiplying the normalization factor
    \begin{equation}
        \label{eq:neff}
        {\cal N}=\int d \omega \left(- \frac{\partial f}{\partial \omega}\right) (\omega-\Delta(\omega))^2
    \end{equation}
    and introducing
    \begin{equation}
        \label{eq:neff}
        \frac{n_\text{eff}}{ m^*}=\frac{{\cal N}}{6\pi^2 \hbar^3 v} .
    \end{equation}
    With
    \begin{equation}
    \label{eq:tau}
    \tau=\hbar \Big \langle \frac{1}{2\Gamma(\omega)} + \frac{3 \Gamma(\omega)}{2 (\omega-\Delta(\omega))^2}   \Big \rangle_{f\omega^2}
    \end{equation}
    we arrive at Equation (6) of the main text. The inclusion of the real part of the self-energy, at variance with ref. \cite{tabert_optical_2016} will change only the definition of the intraband term Eq. (\ref{eq:intra}) which is a  renormalization of the effective number of carriers due to the interaction-induced change of $v$.
    
    We now consider different sources of scattering. The Scattering mechanism affects only the scattering time and not the effective number of carriers which depends only on temperature. From Eq. (\ref{eq:neff}) neglecting the effect of the real part of the self-energy we have
    \begin{equation}
        \frac{n_\text{eff}}{m}= \frac{1}{6\pi^2 \hbar^3 v}  \int d\omega \left(- \frac{\partial f}{\partial \omega}\right) \omega^2.
    \end{equation}
     The integral can be done giving
    \begin{equation}
        \frac{n_\text{eff}}{m}= \frac{T^2}{18 \hbar^3 v} 
    \end{equation}
    which shows a vanishing of effective number of carriers as temperature decreases due to the vanishing of the low energy density of states. 
    Now let $N(\omega)=C\omega^2$ with $C=\frac{1}{2 \pi^2 \hbar^3 v^3}$, then $n_\text{eff}/m = \pi^2 v^2 N(T)/9$.

    \paragraph{Local impurities}: within Self Consistent Born Approximation (SCBA) \begin{equation}
        \Gamma(\omega)=\pi s^2 V N(\omega) 
    \end{equation}
    where $s$ is the strength of disorder local fluctuations, $V$ the volume of the unit cell and $N(\omega)$ the density of states. Assuming that $N(\omega)=C\omega^2$ then $\Gamma(\omega)=\pi s^2 V C \omega^2$ and using Eq. (5) of the main text we can calculate the scattering time as
    \begin{equation}
        \label{eq:taudisorder}
        \tau=\hbar \langle \frac{1}{2\pi s^2 V C\omega^2} + \frac{3 \pi s^2 V C }{2 }   \rangle_{f\omega^2}.
    \end{equation}
    By scaling the temperature in the integrals appearing in Eq. (\ref{eq:taudisorder}) we obtain
    \begin{equation}
        \tau = \hbar \left( \frac{3}{2\pi^3 C s^2 V T^2}+ \frac{3 \pi s^2 V C }{2 }\right).
    \end{equation}
    The second term is negligible for small temperature. Then only (a part of) intraband transitions contribute to the scattering.
    The scattering time diverges as the temperature approaches zero. However taking into account the vanishing of the effective number of carriers
    ($n_\text{eff}/m^*= \pi^2 v^2 N(T)/9$)
    we arrive to a constant residual resistivity
    \begin{equation}
        \sigma =\frac{e^2 v^2 \hbar}{6\pi s^2 V}.
    \end{equation}

    \paragraph{Classical phonons}: this case is similar to that of local disorder if we consider classical phonons (Temperature greater than Debye energy). Then using again SCBA we formally treat phonons as temperature-dependent disorder with variance given by $s^2=2\lambda T D$ where $\lambda$ is the electron-phonon dimensionless coupling constant and $D$ the cutoff. Now $s$ depends on temperature therefore scattering time diverges faster than the vanishing of effective number of carriers leading to
    \begin{equation}
        \sigma =\frac{e^2 v^2 \hbar}{12 \pi \lambda T D V}.
    \end{equation}
    and the resisitivity goes linearly as in the Fermi liquid case.
    
    \paragraph{Long-range interactions}: here $\Gamma(\omega)=\alpha \text{max}(\omega, T)$ \cite{burkov_topological_2011,hosur_charge_2012} with $\alpha$ being a coupling constant. The scattering time obtained by Eq. (\ref{eq:tau}) is thus
    \begin{equation}
    \label{eq:taulr}
        \tau =\hbar \left( \frac{1}{2\alpha T} {\cal I}_1+\frac{3\alpha}{2 T} {\cal I}_2 \right)
    \end{equation}
    where
    \begin{eqnarray}
    {\cal I}_1&=&\frac{3}{4\pi^2} \left (\int_{-1}^{1} dx \frac{x^2}{\cosh^2 (x/2)} + 2\int_1^{\infty} dx  \frac{x}{\cosh^2 (x/2)} \right )\\
    {\cal I}_2&=&\frac{6}{4\pi^2} \left (\int_{0}^{1} dx \frac{1}{\cosh^2 (x/2)} + \int_1^{\infty} dx  \frac{x}{\cosh^2 (x/2)} \right ).
    \end{eqnarray}
    From Eq. (\ref{eq:taulr}) we notice that the temperature behaviour induced in the scattering time by intraband and interband term is the same and gives a linear divergence for low temperature. However the effective number of carrier vanishes more rapidly and the conductivity goes linearly to zero at low temperature producing a divergent resitivity.
    \paragraph{Hubbard interaction}:
        Using $\Gamma(\omega) =\Gamma_0 \left( a_0 \omega^8 +  a_1 \omega^4 T^4 +a_2 \omega^2 T^6 + a_3 T^8\right) $ with $\Gamma_0$ and $a_i$ given in Eq. \eqref{eq:imsigma}  leads to
        \begin{equation}
            \sigma = \frac{e^2}{6\pi^2 \hbar^2 v}  ( \alpha_1 \frac{1}{\Gamma_0 T^6} + \alpha_2 \Gamma_0 T^8).
        \end{equation}
        with $\alpha_1 \approx 0.0032$ and $\alpha_2 \approx 530.91$. At low temperature the second term can be neglected and we arrive at $\sigma\propto T^{-6}$ as discussed in the main text.

\section{vertex corrections}
\label{sec:vtx}
\begin{figure}[th]
\centering
\includegraphics[width=0.3\linewidth]{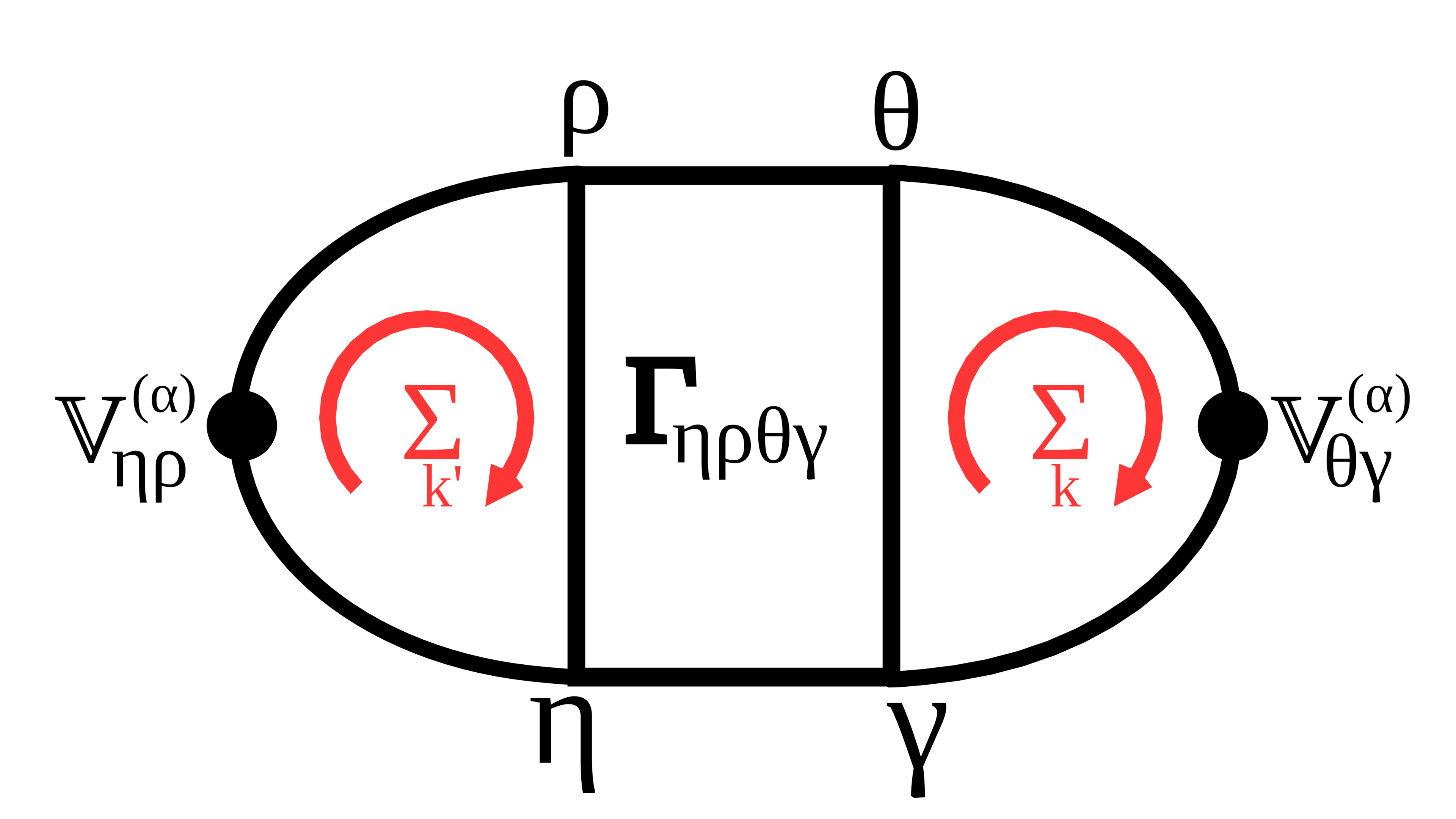}
\hspace{1 cm}
\includegraphics[width=0.23\linewidth]{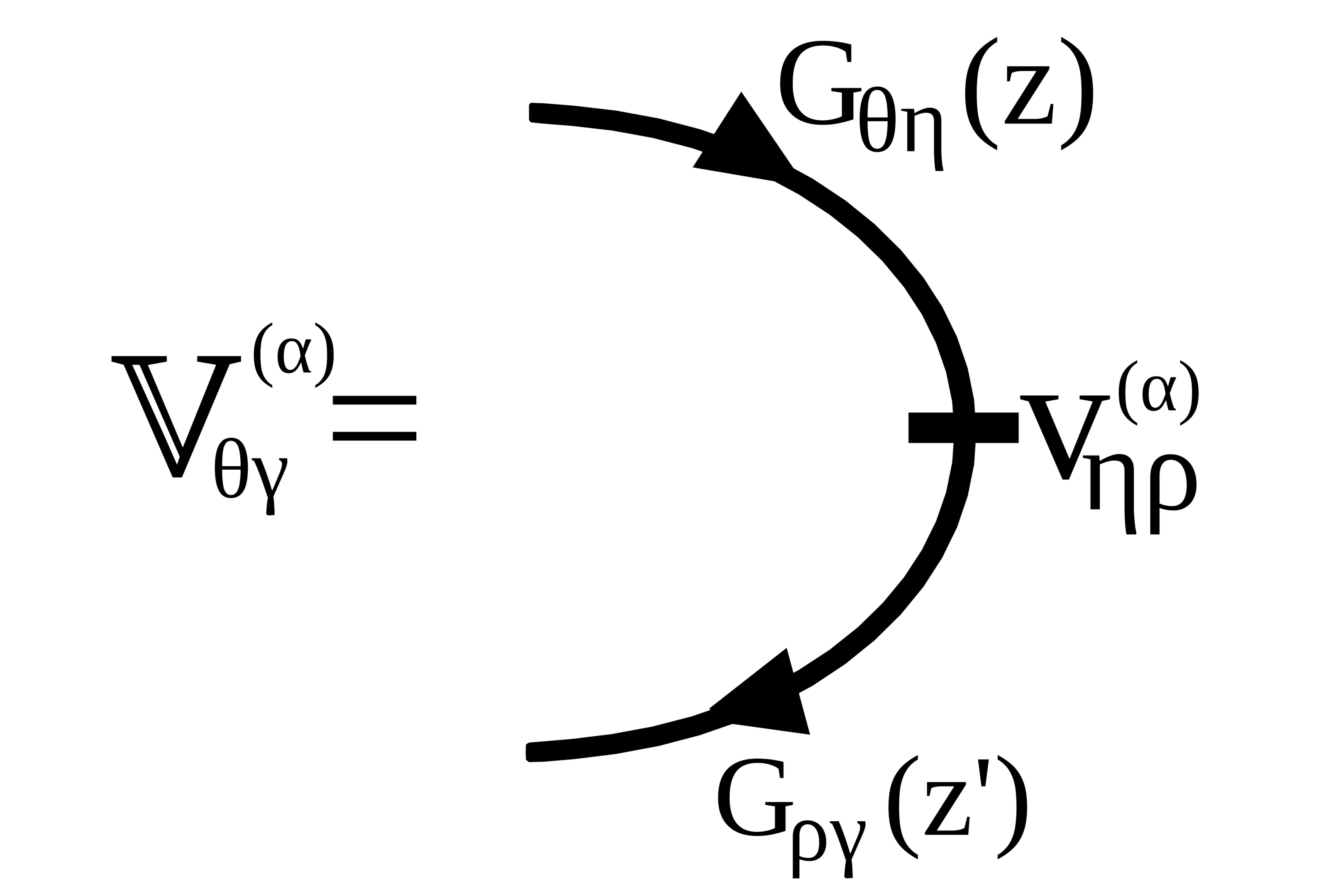}

\caption{Left panel: the conductivity bubble including the vertex $\bm{\Gamma}$. Right panel: definition of $\pmb{\mathbb{V}}$. Here G is the Greens function and $\bf{v}$ the current operator.  }
\label{fig:vtx}
\end{figure}

While vertex corrections to the spin/charge response functions can become very large in DMFT\cite{Uhrig2009,Toschi2012,Galler2015,Hausoel2017,Watzenboeck2020}, their effects on the electrical and thermal conductivities are customarily neglected. Indeed, at the level of {\sl single orbital} calculations in DMFT,  all vertex corrections to the conductivities {\sl exactly} vanish\cite{Khurana1990,georges_dynamical_1996} after performing the internal momentum summations,  as a result of the interplay between the pure locality of the two-particle irreducible vertex $\Gamma$ of DMFT\cite{georges_dynamical_1996,DelRe2020} and the odd parity of the current operator (red loop in Fig.~\ref{fig:vtx}) . 

This symmetry argument applies, however, only to {\sl massive} particles. In the Weyl case, instead, the current operator ${\bf v}^{(\alpha)}=\partial_{k_\alpha} H(\vec{k})$ is a {\sl constant} in momentum space, while it acquires a specific spin structure:
\begin{equation}
\label{eq:currentvtx}
{\bf v}^{(\alpha)} = \sigma_z \otimes \tau_\alpha.
\end{equation}
Here boldface indicates a matrix in the spin and orbital space and $\alpha$ is the direction of the external field.
In order to show that the vertex corrections vanish in the Weyl case we can calculate the diagram appearing in the right panel of Fig. \ref{fig:vtx} i.e.
\begin{equation}
\pmb{\mathbb{V}}^{(\alpha)}(z,z^\prime)=\sum_{{\bf k}} {\bf G}_{\bf k}(z) {\bf v}^{(\alpha)} {\bf G}_{\bf k}(z^\prime).
\label{eq:GvG}
\end{equation}
In Eq. (\ref{eq:GvG}) sum over internal spin indexes, in each orbital sector, is implied.
The Green's function is given by ${\bf G}_{\bf k}(z)= [z {\bf 1} \otimes {\bf 1} - \sigma_z \otimes (\vec{k} \cdot \vec{\tau} )]^{-1} $  or explicitly
\begin{equation}
\label{eq:Green}
{\bf G}_{\bf k}(z)= \frac{1}{z^2 -|k|^2} \left ( z {\bf 1} \otimes {\bf 1}  +\sigma_z \otimes (\vec{k}\cdot \vec{\tau}) \right ).
\end{equation}
In Eqs. (\ref{eq:GvG},\ref{eq:Green}) $z=\omega-\Sigma(\omega)$ with $\Sigma(\omega)$ being the self-energy. Using Eq. (\ref{eq:Green}) in Eq. (\ref{eq:GvG}) we have
\begin{equation}
\pmb{\mathbb{V}}^{(\alpha)}(z,z^\prime)=\sum_{{\bf k}} \frac{1}{z^2 -|k|^2} \frac{1}{z^{\prime 2} -|k|^2} \left ( z {\bf 1} \otimes {\bf 1}  +\sigma_z \otimes (\vec{k}\cdot \vec{\tau}) \right ) \left(\sigma_z \otimes \tau_\alpha \right)  \left (z' {\bf 1} \otimes {\bf 1}  +\sigma_z \otimes (\vec{k}\cdot \vec{\tau})\right ).
\label{eq:GvG1}
\end{equation}
Since ${\bf v}$ is now even in $k$ using Eq. (\ref{eq:currentvtx}) we are left with the following expression
\begin{equation}
\pmb{\mathbb{V}}^{(\alpha)}(z,z^\prime)=\sum_{{\bf k}} \frac{1}{z^2 -|k|^2} \frac{1}{z^{\prime 2} -|k|^2} \left ( z z^\prime \sigma_z \otimes \tau_\alpha +\sigma_z \otimes (\vec{k}\cdot \vec{\sigma}) \tau_\alpha  (\vec{k}\cdot \vec{\sigma}) \right ).
\label{eq:GvG2}
\end{equation}
Using the anticommutation relations for the Pauli matrix we arrive at
\begin{equation}
\pmb{\mathbb{V}}^{(\alpha)}(z,z^\prime)=\sigma_z \otimes \tau_\alpha \sum_{{\bf k}} \frac{z z^\prime-|k|^2}{(z^2 -|k|^2)(z^{\prime 2} -|k|^2)} 
+\sigma_z \otimes \sum_{{\bf k}} \frac{2 k_\alpha (\vec{k}\cdot \vec{\tau})}{(z^2 -|k|^2)(z^{\prime 2} -|k|^2)} .
\label{eq:GvG3}
\end{equation}
The second term in Eq. (\ref{eq:GvG3}) can be further simplified noting that after summation in $k$ only $k_\alpha\tau_\alpha$ term survives leading to
\begin{equation}
\pmb{\mathbb{V}}^{(\alpha)}(z,z^\prime)=\sigma_z \otimes \tau_\alpha \sum_{{\bf k}} \frac{z z^\prime-|k|^2+2k^2_\alpha}{(z^2 -|k|^2)(z^{\prime 2} -|k|^2)} 
\label{eq:GvG4}
\end{equation}
as a consequence the whole spin structure of the diagram is embodied in the Pauli matrix $\tau_\alpha$.\\
Let us consider the simple bubble without vertex corrections:
\begin{equation}
{\cal B}^{(0)(\alpha)}(z,z^\prime)= \operatorname{Tr}{\bf v}_{\alpha}   \pmb{\mathbb{V}}^{(\alpha)}(z,z^\prime) = \operatorname{Tr} \sigma_z \otimes \tau_\alpha   \pmb{\mathbb{V}}^{(\alpha)}(z,z^\prime)
\end{equation}
Inserting the expression for $\pmb{\mathbb{V}}$ from above leads to
\begin{equation}
{\cal B}^{(0)(\alpha)}(z,z^\prime)=4 \sum_{{\bf k}} \frac{z z^\prime-|k|^2+2k^2_i}{(z^2 -|k|^2)(z^{\prime 2} -|k|^2)} .
\label{eq:bubble0}
\end{equation}
Using this equation and  performing the sum over frequency it is possible to derive Eq. (\ref{eq:kubo}).

\begin{figure}[th]
\centering
\includegraphics[width=0.29\linewidth]{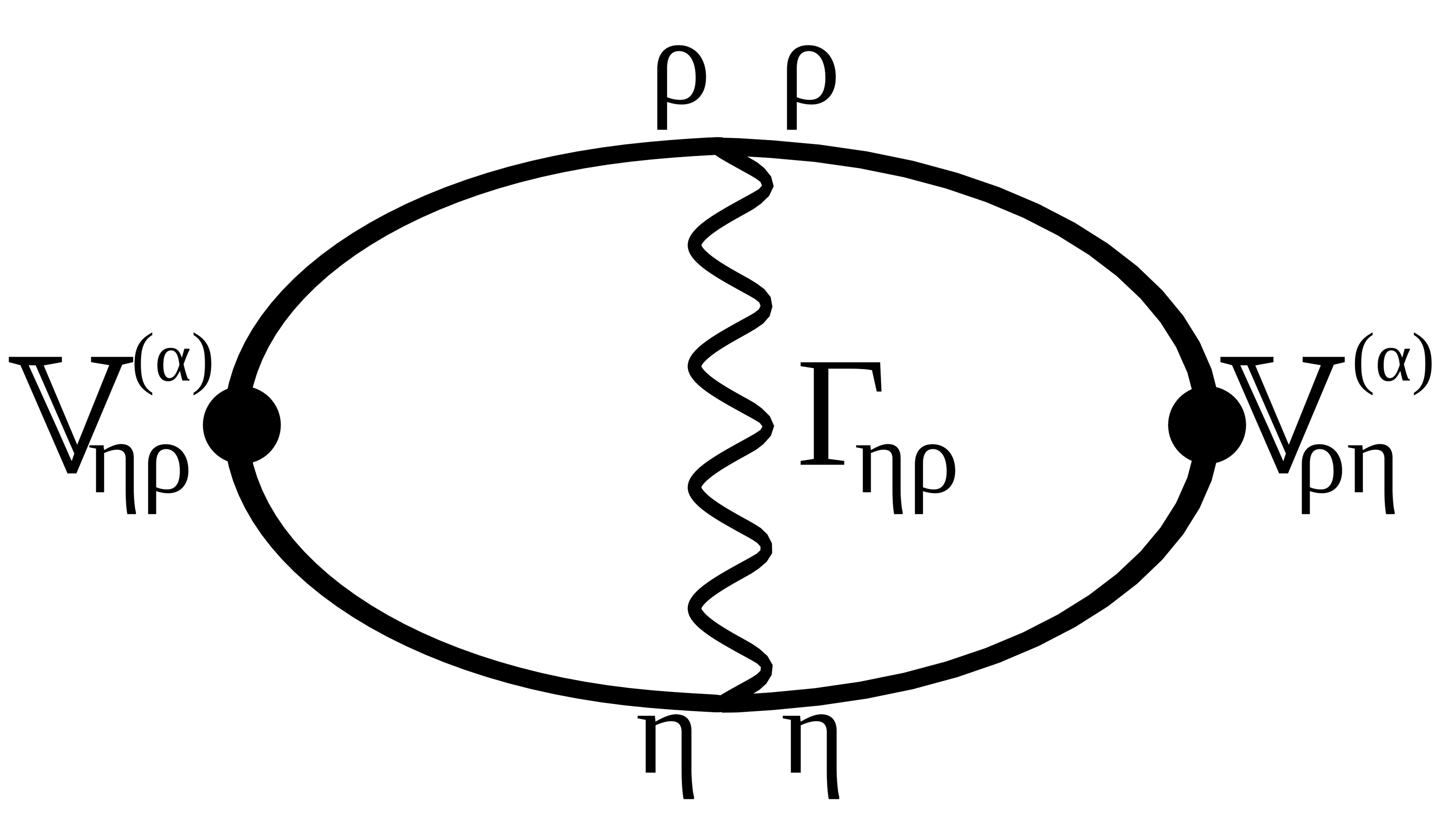}
\hspace{1 cm}
\includegraphics[width=0.3\linewidth]{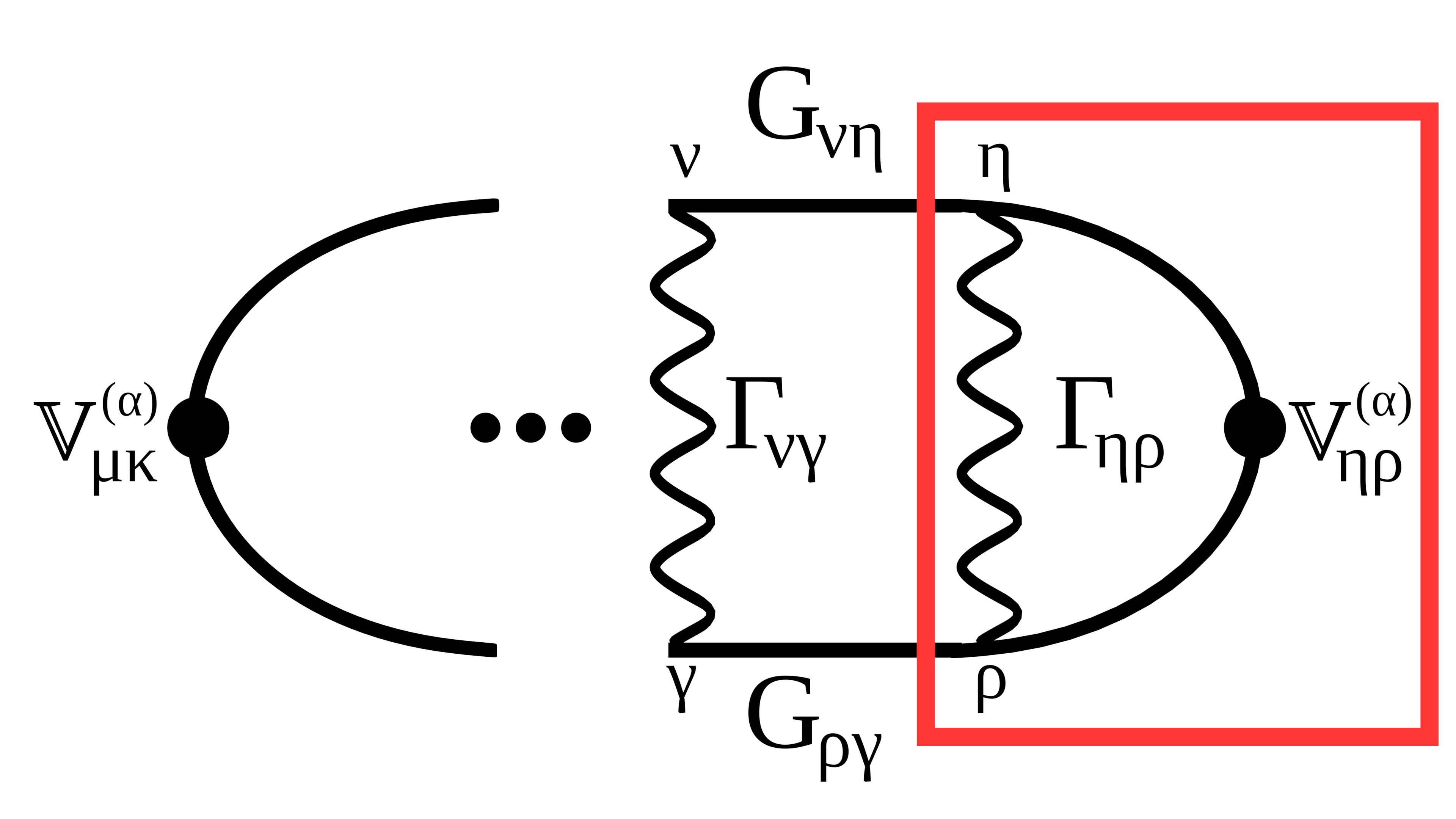}
\caption{Left panel: lowest order vertex correction. Right panel: higher-order ladder diagram.  }
\label{fig:ladder}
\end{figure}

Now let us consider the vertex correction contribution to the conductivity bubble
\begin{equation}
{\cal B}(z,z')= \operatorname{Tr} \pmb{\mathbb{V}}(z,z) {\bf \Gamma} (z,z') \pmb{\mathbb{V}}(z',z')
\label{eq:bubble00}.
\end{equation}
In the previous equation the function $\Gamma$ is local but is a tensor in the four-dimensional spin and orbital space.\\
Let us begin with the lowest-order vertex correction (left panel of Fig. \ref{fig:ladder}). In that case  the vertex $\bm{\Gamma}_{\eta \rho \theta \gamma} =\bm{\Gamma}{\eta \rho \theta \gamma} \delta_{\theta \rho} \delta_{\gamma \eta} =\bm{\Gamma}_{\eta \rho \rho \eta} = \Gamma_{\eta \rho} $ is given by
\begin{equation}
    \Gamma = U \sigma_x  \otimes \tau_0
\end{equation}
because the interaction couples different spin on the same orbital. The bubble including the lowest-order vertex correction is then given by
\begin{equation}
  {\cal B}^{(1)(\alpha)}(z,z') =  \pmb{\mathbb{V}}_{\eta \rho}^{(\alpha)}(z,z)  \Gamma_{\eta \rho} \pmb{\mathbb{V}}_{\rho\eta }^{(\alpha)}(z^\prime,z^\prime).
\end{equation}
When summing over the internal indices this expression is zero for all $\alpha$, demonstrating the vanishing of the lowest-order vertex correction to the bare bubble. Let's recall that this cancellation is a consequence of the tensor structure of the vertex given in Eq.~\ref{eq:currentvtx} and a Hubbard repulsion without inter-orbital contributions.\\
This cancellation is also present for higher-order ladder diagrams (right panel Fig. \ref{fig:ladder})
\begin{equation}
     {\cal B}^{(i)(\alpha)}(z,z') =  \pmb{\mathbb{V}}_{\mu \kappa}^{(\alpha)}(z,z) \dots \Gamma_{\nu \gamma} G_{\nu \rho} G_{\eta \gamma}\cdot \Gamma_{\eta \rho} \pmb{\mathbb{V}}_{\rho\eta }^{(\alpha)}(z^\prime,z^\prime).
\end{equation}
In fact, already the rightmost part of the diagram, i.e. all elements of the matrix $X_{\eta \rho}:=\Gamma_{\eta \rho} \pmb{\mathbb{V}}_{\rho\eta }^{(\alpha)}(z^\prime,z^\prime)= \Gamma_{ \rho \eta} \pmb{\mathbb{V}}_{\rho\eta }^{(\alpha)}(z^\prime,z^\prime)$ vanish before even carrying out the sum over internal indices. This is due to the structure of $\Gamma$ which is non-zero only on the off-diagonal blocks and  $\pmb{\mathbb{V}}$ which is non-zero only on the diagonal blocks.

\section{specific heat}
\begin{figure}[h]
    	\centering
    	\includegraphics[width=0.7\linewidth]{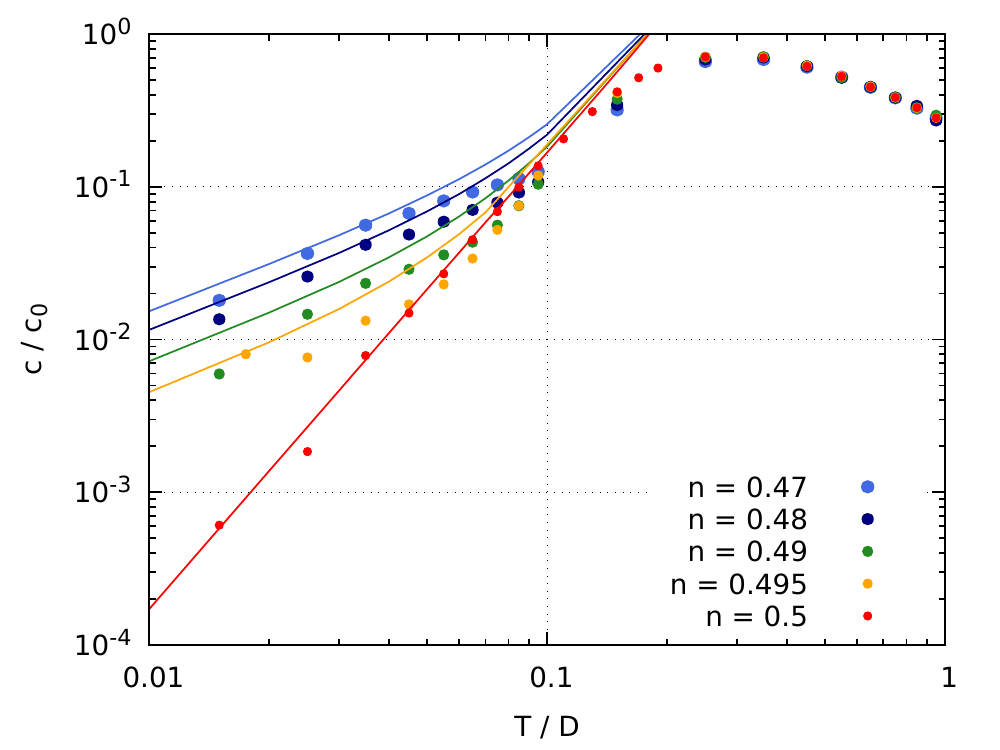}
    	\caption{ DMFT (dots) and analytic (lines) results for the specific heat at different fillings ($U / D=1.0$). $c$ is given in units of $c_0=k_B D^3 /(\hbar v)^3$.}
    	\label{fig:spec_heat_doped}
    \end{figure}

    The specific heat can be computed by evaluating the entropy  \cite{abrikosov_methods_1975,lai_weylkondo_2018}:
    \begin{equation}
        S= \frac{k_B}{2 \pi i T} \int d \epsilon N(\epsilon) \int d \omega \frac{\partial f}{\partial \omega}  \omega [ \log G_R(\epsilon,\omega) -\log G_A(\epsilon,\omega) ].
    \end{equation}
    \begin{equation}
    c_v(T)= k_B\frac{T}{Z} \int dy d \epsilon N(\epsilon) \frac{y^2 e^y}{(e^y+1)^2} A(\epsilon,y T).
    \end{equation}
    with $A(\epsilon,\omega)= -\frac{1}{\pi } \operatorname{Im} G_R(\epsilon,\omega) = Z \delta(\omega-Z \epsilon)$. Here $Z$ is the quasiparticle weight. This leads to:
    \begin{equation}
    c_v(T)=k_B \frac{T}{Z} \int dy N(y T/Z) \frac{y^2 e^y}{(e^y+1)^2} .
    \end{equation}
    For a quadratic dispersion $N(\omega) = \frac{3}{2 (\hbar v)^3} \omega^2 $ this yields (including a factor of two for the two spin configurations):
    \begin{equation}
    c_v(T) = k_B \frac{1}{ (\hbar v)^3} \frac{7 \pi^4}{5} \frac{T^3}{Z^3} .
    \end{equation}
    Doping away from the degeneracy point, i.e. $N(\omega) = \frac{3}{2 (\hbar v)^3} (\omega-\mu)^2$, leads to
    \begin{equation}
        c_v(T) =k_B \frac{1}{(\hbar v)^3}  \frac{7 \pi^4 T^3 + 5 \pi^2 Z^2 T \mu^2}{5 Z^3} .
    \end{equation}
    Figure \ref{fig:spec_heat_doped} displays DMFT results compared to the analytic formula for different fillings. The results show qualitative agreement. The differences can be attributed to the fact that the spectral function of the DMFT results differs slightly from the parabolic form which enters the analytic calculations.
    \clearpage
	
\section{Relation to real materials}
The temperature range where $T^\alpha$ with $\alpha>2$ can be expected is restricted from below by the chemical potential, i.e. the distance of the degeneracy point from the Fermi level, and from above by the 'effective cut-off' $D_{\text{eff}}$, i.e. the extension of the linear dispersion in the band structure. For Cd$_2$As$_3$ the latter can be estimated using a summary of different experimental and theoretical works presented in Ref.~\onlinecite{crassee_3d_2018}. There, an energy scale $E_D$ is given for the Dirac cones which ranges from $20\,meV$ to several hundred $meV$. Because the dispersion starts to deviate from linear behaviour already before $E_D$ is reached, we can use $E_D$ as an upper bound for the effective cutoff. Using the smallest reported value $E_D =20\,meV$ the upper bound in temperature: $k_B T = 0.1 D_{\text{eff}}$ can be estimated to be $23\,K$ which is of the same order as the transition shown in Fig. \ref{fig:exp_data}. The transition from $T^2$ to $T^\alpha$ with $\alpha>2$ takes place at $T\approx9\,K$ which using $k_B T=0.2 \mu_{\text{eff}}$ for the transition temperature yields $\mu_{\text{eff}} \approx 4\,meV$.\\
Note that the temperature dependence of the resistivity of Cd$_2$As$_3$ varies strongly between different samples measured in Ref.~\onlinecite{liang_ultrahigh_2015}. Not all samples show the $\alpha>2$ behaviour.
This suggests that the exponent is strongly sensitive to sample preparation and exact position of the chemical potential.\\
Figure \ref{fig:exp_data} also displays data for MoP \cite{kumar_extremely_2019} and WP$_2$ \cite{kumar_extremely_2017} which both show qualitative similar behaviour when compared to Cd$_2$As$_3$.  
\begin{figure*}[h]
\centering
	\includegraphics[width=0.5\linewidth]{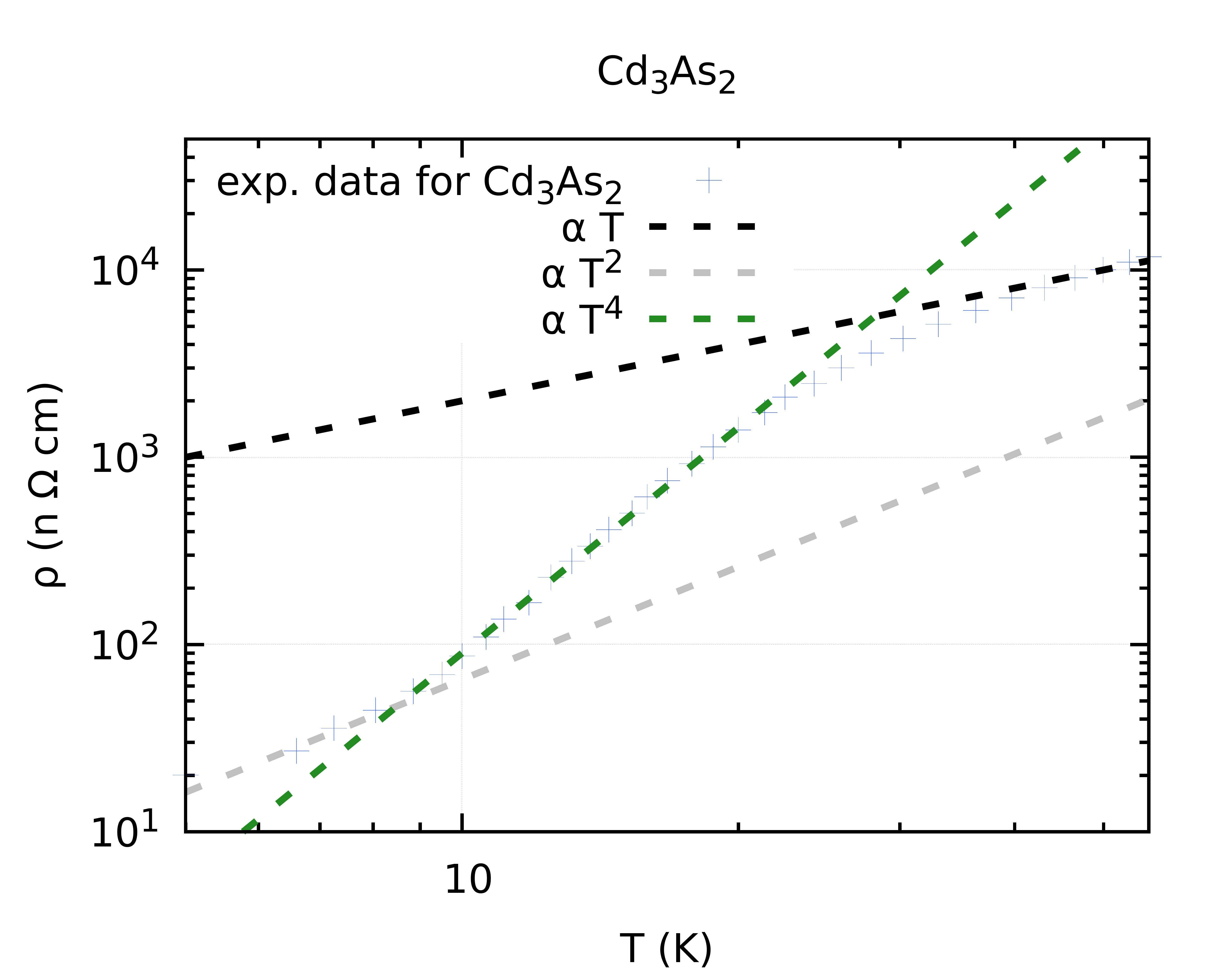}
\end{figure*}
\begin{figure*}[h]
\centering
	\includegraphics[width=0.5\linewidth]{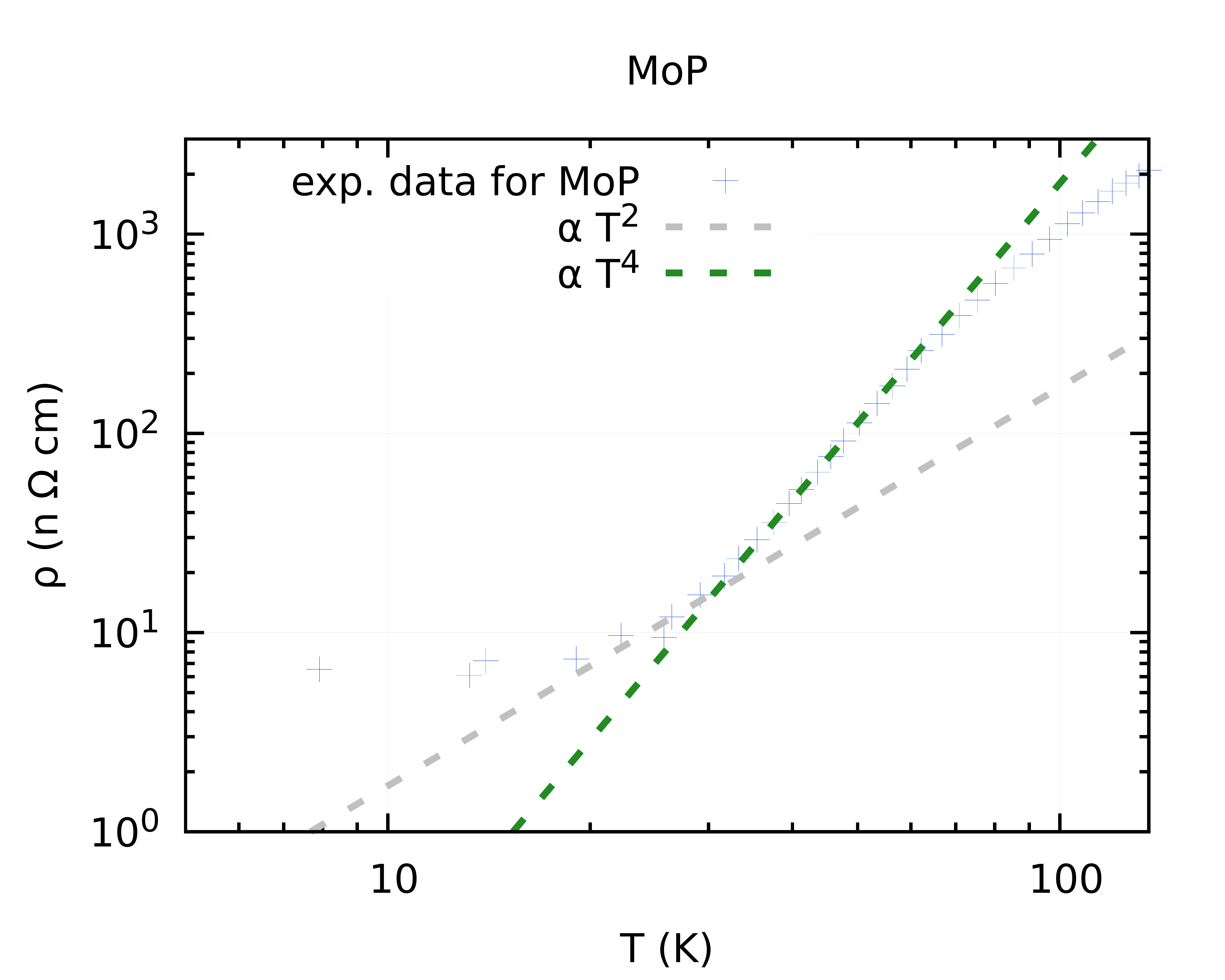}
\end{figure*}
\begin{figure}[h]
	\centering
	\includegraphics[width=0.5\linewidth]{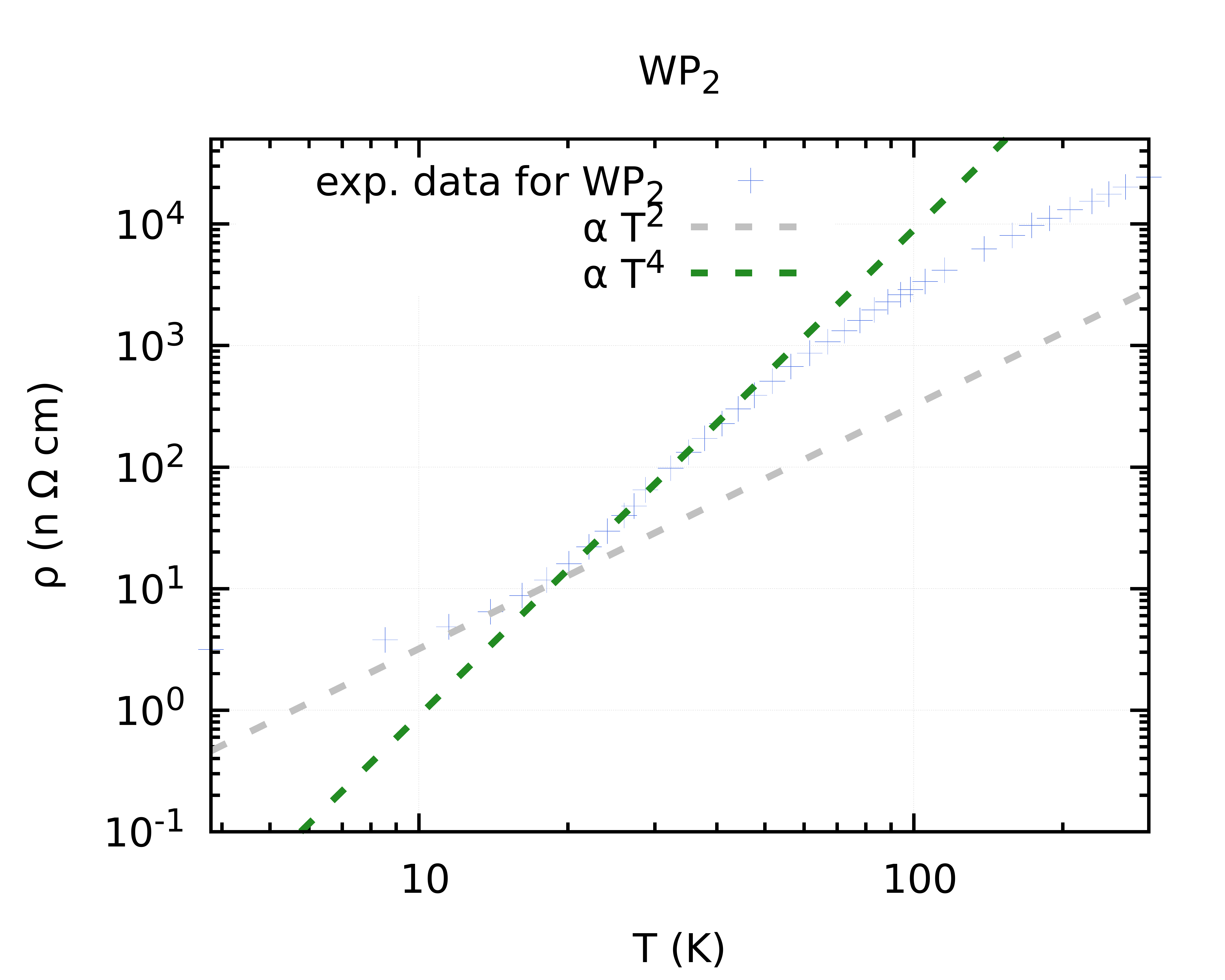}
	\caption{Experimental results for the resistivity as a function of temperature of Cd$_2$As$_3$ (taken from Ref. \onlinecite{liang_ultrahigh_2015}), MoP (taken from Ref. \onlinecite{kumar_extremely_2019}) and WP$_2$ (taken from Ref. \onlinecite{kumar_extremely_2017}).}
	\label{fig:exp_data}
\end{figure}
\clearpage
\bibliography{references.bib}